\documentclass[%
aps,
prl,%
amsmath,amssymb,showpacs,
reprint,%
floatfix,
twocolumn,
nofootinbib,
longbibliography,
superscriptaddress
]{revtex4-2}

\usepackage[dvipsnames]{xcolor}

\usepackage[disable]{todonotes}
\usepackage{amsmath,amsfonts,amssymb}
\usepackage{cases} 
\usepackage{xfrac}

\usepackage{graphicx}

\usepackage{hyperref}

\allowdisplaybreaks[3]

\newcommand{\newt}[1]{{\color{black}#1}}

\DeclareMathAlphabet\mathbfcal{OMS}{cmsy}{b}{n}

\usepackage{xr}
\newcommand{\figLettInv}{4}

\begin{document}
	
	\title{Towards nonlinear optics with M\"ossbauer nuclei using x-ray cavities}
	
	\author{Dominik \surname{Lentrodt}}
	\email[]{dominik.lentrodt@physik.uni-freiburg.de}
	
	\affiliation{Physikalisches Institut, Albert-Ludwigs-Universit\"at Freiburg, Hermann-Herder-Stra{\ss}e 3, D-79104 Freiburg, Germany}
	\affiliation{EUCOR Centre for Quantum Science and Quantum Computing, Albert-Ludwigs-Universit\"at Freiburg, Hermann-Herder-Stra{\ss}e 3, D-79104 Freiburg, Germany}
	\affiliation{Max-Planck-Institut f\"ur Kernphysik, Saupfercheckweg 1, 69117 Heidelberg, Germany}
	
	\author{Christoph H.~\surname{Keitel}}
	
	\author{J\"org \surname{Evers}}
	\email[]{joerg.evers@mpi-hd.mpg.de}
	
	\affiliation{Max-Planck-Institut f\"ur Kernphysik, Saupfercheckweg 1, 69117 Heidelberg, Germany}
	
	\begin{abstract}
		Strong excitation of nuclear resonances, particularly of Mössbauer nuclei, has been a longstanding goal and the advance of novel x-ray sources is promising new options in this regard. Here we map out the necessary experimental conditions for the more general goal of realizing nonlinear optics with nuclei and compare with available technology. In particular, we present a comprehensive theory of nonlinear nuclear excitation in thin-film x-ray cavities by focused x-ray pulses. We thereby identify cavity geometries with broad resonances which allow one to boost the nuclear excitation even at moderately tight focusing \newt{and offer the possibility to mitigate radiation damage}.
	\end{abstract}
	
	\maketitle

	In this paper, we revisit an old challenge: Is it possible to fully excite an ensemble of atomic nuclei using externally applied electromagnetic fields? This question relates to the goal of building a $\gamma$-ray laser (``graser''), with inverted nuclei as a gain medium, which was proposed long ago after the laser-maser principle~\cite{Baldwin1981}. The discovery of the M\"ossbauer effect, which allows for the recoil-less absorption and emission of $\gamma$-ray photons from certain nuclei \cite{Moessbauer1958,Moessbauer1962}, further fueled this research. Despite interim hope to circumvent difficulties by utilizing lasing without inversion \cite{Kocharovskaya1992}, most approaches to grasing were deemed unfeasible by the late 1990s~\cite{Baldwin1997}. Nevertheless, recent technological and conceptual developments prompt us to reconsider the challenge.
	
	First, advances in x-ray source technology such as x-ray free-electron lasers (XFEL)~\cite{Pellegrini2016,Georgescu2020} have promoted the study of high-energy nonlinear effects in atoms, analogous to the revolutionary development at lower photon energies following  the invention of conventional laser sources~\cite{Schawlow1958,Lamb1999}. Various nonlinear x-ray processes have already been observed involving electronic transitions~\cite{Matsuda2023,Rohringer2019,Sofer2019,Nandi2022,Doumy2011}. Closely related to the original question, also lasing on an inner-shell electronic transition based on XFEL-generated inversion has been demonstrated~\cite{Rohringer2012}. This raises the question if the source advances provide new avenues for the excitation of nuclei.
	
	Second, M\"ossbauer nuclei by now have evolved into a promising platform for quantum optics at energies of hard x-rays. In a series of experiments, key concepts of quantum optics~\cite{Rohlsberger2010,Rohlsberger2012,Heeg2013a,Heeg2015a,Heeg2015b,Haber2016a,Haber2017} and coherent control~\cite{Helisto1991,Shvydko1996,Schindelmann2002,Vagizov2014,Sakshath2017,Heeg2017,Heeg2021,Bocklage2021} have been demonstrated with nuclei as the matter part. However, while these experiments exploit coherence  and interference effects and demonstrate a surprising level of control, they still operate in the linear optics regime, such that they can be described by (semi-)classical optics calculations \cite{Rohlsberger2005,Sturhahn2000}.
	
	These two developments also motivate a shift in focus away from the goal of reaching full inversion and towards more general nonlinear optics effects, further fueled by recent experiments from the optical domain \cite{Manzoni2017,Mahmoodian2018,Prasad2020,Cordier2023,Liedl2024} demonstrating that effects such as photon correlations can already appear far below full excitation. First experiments with M\"ossbauer nuclei at XFELs~\cite{Chumakov2018,Shvydko2023} explore this direction, e.g.~by scattering multiple photons per shot \cite{Chumakov2018}. Still these experiments do not reach beyond linear optics, which remains an open challenge.
	
	On the theoretical side, a number of schemes have been proposed to achieve higher excitation or get close to the inversion threshold. For example, it has been suggested to overcome discouraging first estimates~\cite{Matinyan1998} using additional accelerations of the target nuclei~\cite{Buervenich2006}, non-dipole transitions~\cite{Palffy2008}, or collective effects in the light-matter interaction~\cite{HeegPhD,Junker2012}. Furthermore, there are multiple approaches to boost the excitation of the nuclei by a suitable tailoring of their electromagnetic environment. As one option, front-coupling in one-dimensional x-ray waveguides is well-studied for imaging applications~\cite{Salditt2020,Schenk2011PhD}, has recently been implemented with nuclei \cite{Lohse2024_arxiv,Lohse2024a}, and tapered variants have been suggested as a candidate for nuclear inversion~\cite{Chen2015,Chen2022}. Alternatively, x-ray thin-film planar cavities containing layers of M\"ossbauer nuclei are well established as a major platform for nuclear quantum optics in the linear regime~\cite{Rohlsberger2014,Rohlsberger2021}. The field enhancement inside such structures has been considered \cite{Rohlsberger2005,Rohlsberger2005a,Rohlsberger2010,Heeg2013b,Lentrodt2020a} and optimized~\cite{Schenk2011PhD,Diekmann2022}. However, the latter works consider the state-of-the-art probing using highly collimated x-rays, which is unfavorable for strong excitation. Focusing to enhance the nuclear excitation has been proposed~\cite{Heeg2016arxiv}, but so far, rigorous modelling of the excitation dynamics in planar waveguides driven by focused x-ray beams and a corresponding roadmap towards nonlinear excitation are missing.
	
\begin{figure}[t]
		\centering
		\includegraphics[trim={5.0cm 11.0cm 10.0cm 3.0cm},clip,width=1.0\columnwidth]{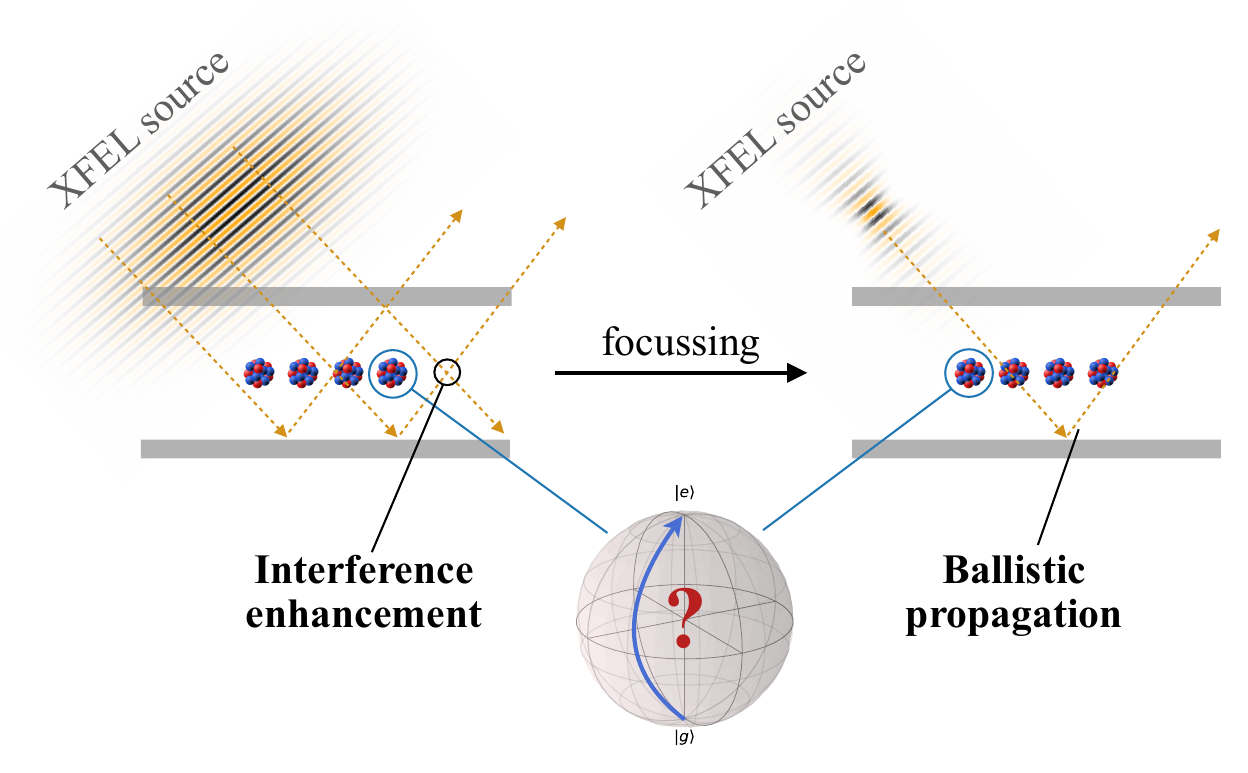}
		\caption{Sketch of the setup, featuring a focused x-ray pulse (not to scale) incident on a thin-film cavity doped with narrow resonances, such as M\"ossbauer nuclei. By focusing the incident x-ray pulse, one can concentrate the radiation on fewer nuclei, potentially increasing their excitation towards inversion (illustrated by the Bloch sphere). However, this comes at the cost of reducing the cavity enhancement from interference. We show that using suitably designed cavities, this limitation can be overcome in intermediate focusing regimes and map out which configurations are most advantageous for upcoming experiments.}
		\label{fig::examples_illu_setup}
	\end{figure}

	In this Letter, we utilize a comprehensive theory developed  in a companion paper \cite{companion_nonlin2024_arxivMethod} to study the excitation dynamics of M\"ossbauer nuclei in planar cavities driven by focused x-ray pulses (see Fig.~\ref{fig::examples_illu_setup}). We find that the rigorous modeling of the focusing leads to optimum cavity structures which qualitatively differ from well-established design paradigms for collimated x-ray beams. The latter may even lead to an excitation performance worse than without the cavity. With varying size of the x-ray focus, we identify a transition between standing-wave interference and ballistic reflections, with significant field enhancement down to moderately tight focusing. We calculate the necessary source parameters to enter the nonlinear regime of significantly excited nuclei and demonstrate its feasibility by comparing with currently available conditions and planned upgrades.
	
	We start with the nuclear excitation dynamics. Unlike in the low-excitation regime~\cite{Asenjo-Garcia2017a,Lentrodt2020a}, the modeling of dissipative quantum dynamics of many interacting few-level systems at higher excitation remains a challenging problem of much current interest~\cite{Mink2023,Schachenmayer2015,Moreno-Cardoner2021}. However, x-ray pulses delivered by accelerator-based  sources are orders of magnitude shorter (typically ps to fs scale) than any other relevant time scale in the nuclear dynamics (e.g., ns  for ${}^{57}$Fe). This allows one to separate the excitation from the subsequent decay dynamics. The former then essentially reduces to a single-particle problem (see~\cite{companion_nonlin2024_arxivMethod} for details), since dissipative and coupling processes are ineffective over the short excitation time scales. 
	
	The single-nucleus dynamics can straightforwardly be solved numerically for arbitrary input fields. However, to gain analytical insight, we further assume resonant and Fourier-limited x-ray pulses, an idealization of pulses generated via hard x-ray self-seeding~\cite{Amann2012,Nam2021,Liu2023}. Then, the pulse area theorem~\cite{McCall1967,Eberly1998,Shore2011} relates the nuclear state after the x-ray pulse to the Fourier transform of the x-ray field at the nuclear resonance frequency $\textbf{E}_\mathrm{drive}(\mathbf{r}_\mathrm{nuc}, \omega_\mathrm{nuc})$ as~\cite{Allen1987_BOOK,companion_nonlin2024_arxivMethod}
	\begin{align}\label{eq::pulse_area_tot_LETT}
		\Phi = \frac{4\pi}{\hbar}|\textbf{d}\cdot\textbf{E}_\mathrm{drive}(\mathbf{r}_\mathrm{nuc}, \omega_\mathrm{nuc})| \,.
	\end{align}
	Here, $\Phi$ is the so-called pulse area, and $\textbf{d}$, $\mathbf{r}_\mathrm{nuc}$, and $\omega_\mathrm{nuc}$ are the electric dipole moment (an effective quantity for magnetic transitions), the position and the transition frequency of the nuclei, respectively. The nuclear excitation --- which is directly related to the magnitude of many nonlinear processes --- is then obtained as ${\sigma}^z = -\cos(\Phi)$, where ${\sigma}^z$ is the expectation value of the transition's Pauli operator.
	
	An intuitive interpretation is gained by rewriting Eq.~\eqref{eq::pulse_area_tot_LETT}~\cite{companion_nonlin2024_arxivMethod} as
	\begin{align}
		\Phi = \frac{\chi_\sigma}{w_0} \cdot \chi_\mathrm{source} \cdot \xi_\mathcal{E} \,,
	\end{align}
	where $\chi_\sigma$ is a transition characteristic that solely depends on the nuclear properties such as the dipole moment. $w_0$ is the beam waist size in the focal plane, characterizing the strength of the focusing. The incident x-ray pulses are characterized by their pulse energy $E_{\mathrm{pulse}}$ and  relative bandwidth $b_r$  via $\chi_\mathrm{source}:=\sqrt{E_\mathrm{pulse}/b_r}$. Finally,  $\xi_\mathcal{E}$ is an enhancement factor due to the cavity. Without cavity --- that is in free space --- it is equal to unity. Therefore, the goal of the cavity environment is to increase $\xi_\mathcal{E}$. From the criterion $\Phi=\pi$ for reaching inversion, we define a \textit{necessary} source characteristic for a given nucleus, cavity and focusing as
	\begin{align}
		{\chi}_\mathrm{source}^\mathrm{nec} = \pi \frac{w_0}{{\chi}_\sigma \,\xi_\mathcal{E}} \,.
	\end{align}
	For a given $\chi_\mathrm{source}$, one can then express the achievable nuclear excitation as $\sigma^z = -\cos\left( \frac{\pi \chi_\mathrm{source}}{{\chi}_\mathrm{source}^\mathrm{nec}}\right)$.
	
	The crucial position-dependent enhancement factor $\xi_\mathcal{E}(\mathbf{r})$ is obtained by calculating the propagation of  focused x-ray pulses in the cavity medium using Maxwell's equations. In the companion paper~\cite{companion_nonlin2024_arxivMethod}, an efficient semi-analytical algorithm specialized for this purpose is developed, which forms the basis of the results below.
	
\begin{figure}[t]
		\centering
		\includegraphics[width=0.9\columnwidth]{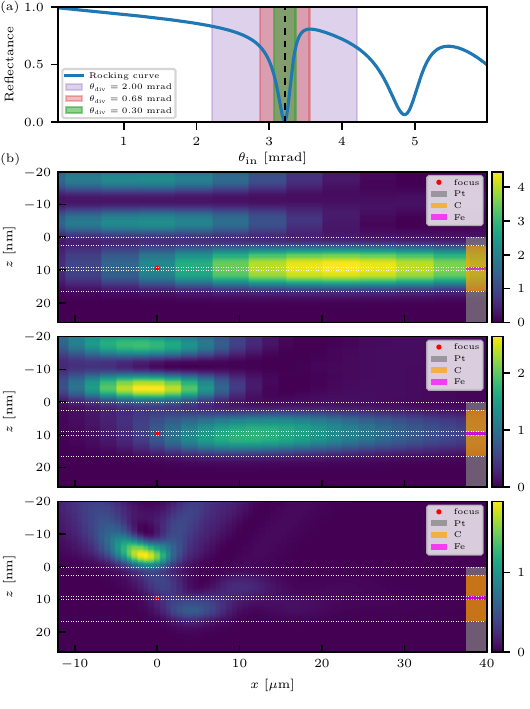}
		\caption{Results for an example cavity following standard design criteria.
			(a) Electronic reflection as a function of incidence angle (rocking curve). The shaded areas indicate three different input beam divergence ranges considered in panel (b), see legend. (b) Spatially resolved cavity enhancement factor $\xi_\mathcal{E}$ in and around the example cavity (cladding boundary at $z=0$) for the beam divergences $\theta_\mathrm{div}=0.3,0.685$ and $2.0\,$mrad (top to bottom), where the middle value corresponds to 40$\,$nm spot size. The incident beam propagates from the top left to the bottom right and encounters the cavity starting from $z=0$. The colored areas and white horizontal lines indicate the location of the cavity layers (see legend). The red dot indicates the free-space beam focus, which is placed at the depth of the nuclear ensemble. Intensities are normalized to corresponding free-space intensity maxima.}
		\label{fig::examples_realisticBeamDivs}
	\end{figure}

	We start with the discussion of a standard cavity design~\cite{supplement_nonlin2024,Rohlsberger2010}. Results are shown in Fig.~\ref{fig::examples_realisticBeamDivs}. Panel (a) depicts the electronic reflectance as a function of grazing incidence angle (rocking curve), featuring two cavity resonances. Following standard design criteria, the first resonance is relatively narrow and critically coupled~\cite{Rohlsberger2010,Schenk2011PhD,Diekmann2022}. The latter condition implies a vanishing of the resonant electronic reflectance, such that the isolated response of the nuclei inside the cavity can experimentally be accessed~\cite{Rohlsberger2002}.

	Panel (b) shows the spatially-resolved cavity enhancement factor inside and outside the cavity for three different x-ray input fields (top to bottom). Their incidence angles are all on resonance with the first cavity mode, but they feature different focusing strenghts. The correponding beam divergences $\theta_\mathrm{div}=0.3,0.685$ and $2.0\,$mrad are indicated as shaded regions in (a).  $x$ is the projection of the pulse propagation direction onto the waveguide surface, and $z$ the surface normal. 
	
	\begin{figure}[t]
		\centering
		\includegraphics[width=0.9\columnwidth]{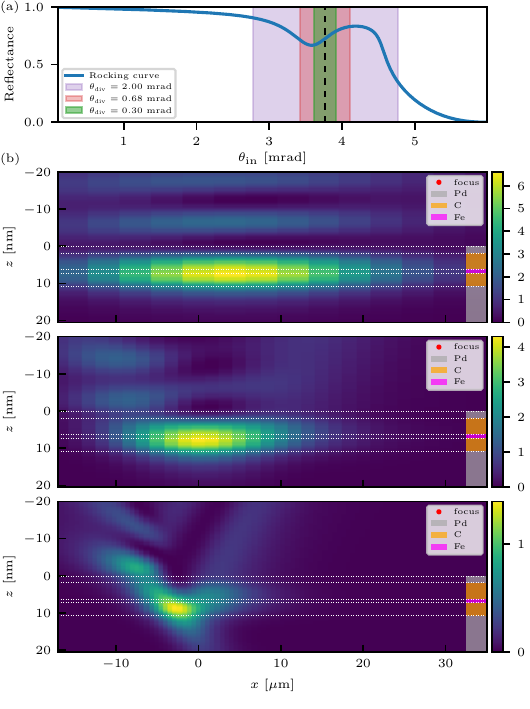}
		\caption{Analogous to Fig.~\ref{fig::examples_realisticBeamDivs}, but for a cavity optimized for a focal spot size of 40~nm. The resonance structure in this case is rather different from the standard cavity, with a broad resonance far from critical coupling. However, such a cavity allows to obtain field enhancement at the resonant ensemble location (magenta) even for tight focusing. Note that the free-space focus is located above the plot limits at 33.2$\,$nm as a result of the optimization.}
		\label{fig::examples_realisticBeamDivs2}
	\end{figure}

	\begin{figure*}[t]
		\centering
		\includegraphics[width=1.0\textwidth]{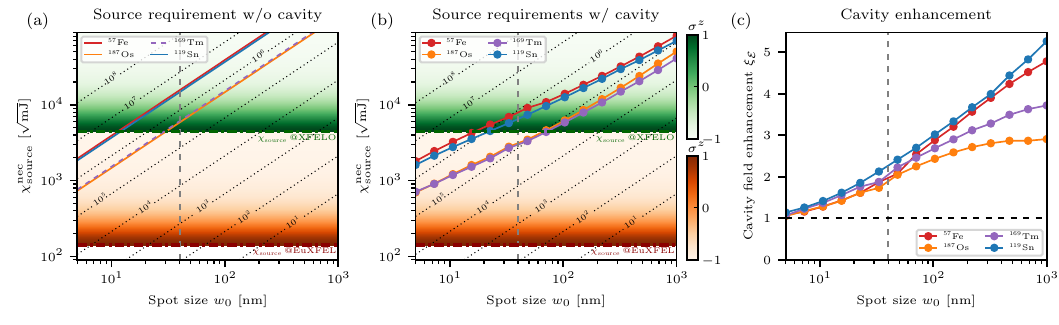}
		\caption{\newt{Source requirements necessary for inversion as a function of the beam spot size, expressed as the necessary source characteristic $\chi^\mathrm{nec}_\mathrm{source}=\sqrt{E^\mathrm{nec}_\mathrm{pulse}/b^\mathrm{nec}_r}$. Panel (a) shows results for each isotope (lines; see legend) in free space. Panel (b) shows corresponding results with a cavity environment optimized for enhancement at each spot size (optimization points marked by dots; lines provided as a guide for the eye). The vertical dashed line indicates a potentially achievable spot size of 40$\,$nm. Panel (c) shows the corresponding cavity enhancement factor. For comparison to practical conditions, the available $\chi_\mathrm{source}$ at the European XFEL (horizontal red dash-dotted line) and at the planned oscillator upgrade \cite{supplement_nonlin2024} (horizontal green dash-dotted line) are shown in (a) and (b). Inversion is then realized when $\chi_\mathrm{source}$ and $\chi_\mathrm{source}^\mathrm{nec}$ cross. As additional information, the background shading (red and green, respectively) indicates the achievable excitation if $\chi_\mathrm{source}$ lies below $\chi_\mathrm{source}^\mathrm{nec}$, that is for sub-inversion conditions. The color axis then corresponds to $\sigma^z=-\cos\left( \frac{\pi \chi_\mathrm{source}}{{\chi}_\mathrm{source}^\mathrm{nec}}\right)$ [colorbars in panel (b)] at a given necessary [${\chi}_\mathrm{source}^\mathrm{nec}$] and actual [$\chi_\mathrm{source}$] source parameter for the respective sources \cite{supplement_nonlin2024}. The physically relevant value is then the shading at the point of the corresponding $\chi_\mathrm{source}^\mathrm{nec}$-line and indicates the expected strength of nonlinear effects which are achievable with each M\"ossbauer isotope (see legend) at this source. Similarly, the diagonal dotted lines indicate contours of x-ray fluences per pulse bandwidth at the cavity surface [units of $\frac{\mu \mathrm{J}}{\mu\mathrm{m}^2\, \mathrm{meV}}$], which allow to estimate the radiation damage on the targets.}}
		\label{fig::inversion_spotSizeScan1}
	\end{figure*}

	At low focusing (panel b, top), the field enhancement inside the cavity is clearly visible, arising from the constructive interference of multiple x-ray reflections.
	As desired, the intensity maximum is at the position of the nuclei.  
	Above the cavity surface, a standing wave  pattern appears from the interference of in- and outgoing x-ray fields. 

	Towards higher focusing [panel (b), top to bottom]  the intensity enhancement reduces. On the one hand, with increasing beam divergence, a growing part of the incident wave vector components is off-resonant and cannot contribute to the standing wave enhancement. Also, however, the focused wave packets become smaller such that the spatial overlap of the different reflections diminishes, again impeding the constructive interference leading to the intensity enhancement. The latter effect is visible in the highly focused case (b, bottom), and implies a practical lower spot-size limit for the focusing. In addition to the enhancement reduction, most of the wave packet is reflected already at the outer cladding, such that the maximum field intensity is shifted to the outside of the cavity structure. Thus, the intensity at the nuclei's positions is in fact lower than it would be in the absence of the cavity. 
	
	From these results, we conclude that standard cavity designs are unsuitable in combination with focusing. However, these limitations can be overcome using alternative cavity designs. To this end, we choose a moderately strong focusing spot size of $40$~nm and optimized the cavity layer structure for maximum intensity enhancement at the position of the nuclei~\cite{supplement_nonlin2024}. The results in this case are shown in Fig.~\ref{fig::examples_realisticBeamDivs2}. In comparison to the standard cavity in Fig.~\ref{fig::examples_realisticBeamDivs}, we observe two qualitative differences. First, the  rocking curve in panel (a) features a relevant resonance which is rather weak and surprisingly broad. Second, the optimized cavity [structure see Fig.~\ref{fig::examples_realisticBeamDivs2}, panel (a)] enables a significant intensity enhancement even at higher focusing [panel (b), top to bottom], and the x-rays are mostly reflected from the substrate rather than from the surface, which boosts the excitation at the nuclear position. This improvement is due to the  larger acceptance angle of broader cavity resonances.
	
	After discussing the qualitative aspects of focusing on the nuclear excitation dynamics, we now turn to quantitative predictions for the M\"ossbauer isotopes  $^{57}$Fe,  $^{187}$Os, $^{169}$Tm and $^{119}$Sn. To this end, we determine the necessary source parameter  ${\chi}_\mathrm{source}^\mathrm{nec}$ to reach nuclear inversion as a function of the free-space beam spot size $w_0$ for each of these isotopes. For comparison, we further indicate the source conditions available at the European XFEL~\cite{Tschentscher2017,Madsen2021,Madsen2013_designRep,Nakatsutsumi2014_designRep} and the projected conditions at the planned oscillator upgrade \cite{Adams2019} (detailed data tables are provided in~\cite{companion_nonlin2024_arxivMethod}). 
	
	Results are shown in Fig.~\ref{fig::inversion_spotSizeScan1}.
	As a reference, panel (a) illustrates the case without cavity environment, for which ${\chi}_\mathrm{source}^\mathrm{nec} \propto \omega_\mathrm{nuc}$.
	In the relevant hard x-ray energy range, nano-focusing setups at EuXFEL beamlines currently offer spot sizes down to about 40$\,$nm~\cite{website_MID_calls}, which is indicated as a vertical dashed line.
	At  this focusing strength, present-day beam conditions at EuXFEL are still far below the requirements for full inversion. However, XFELO conditions~\cite{Adams2019} already approach the required $\chi_\mathrm{source}$ values and significant $\sigma^z$ can be reached, indicating the onset of the nonlinear regime.
	
	Panel (b) shows corresponding results with waveguide enhancements. To this end, we optimized the cavity structure for maximum excitation enhancement \newt{at each spot size} using the algorithm developed in the companion paper~\cite{companion_nonlin2024_arxivMethod}. We see that the cavity enhancement indeed allows one to reduce the source requirements, such that the projected XFELO parameters even exceed the requirements for inversion in  $^{187}$Os and $^{169}$Tm. The relative improvement by the cavity is quantified in panel (c), which shows the ratios of the required source parameters with and without cavity as a function of focusing spot size, which is equal to the cavity field enhancement $\xi_\mathcal{E}$.  We see that enhancements in the pulse area are possible even at substantial focusing. Note that away from full inversion, which presently is experimentally most relevant, the excitation scales quadratically with the pulse area, such that nonlinear effects are also enhanced quadratically.
	
	\newt{In addition to the available source and focusing conditions, an important practical question is whether the target can survive possible radiation damage. To this end, Fig.~\ref{fig::inversion_spotSizeScan1} also shows the fluences per pulse bandwidth on the sample surface. Damage thresholds of materials in the x-ray regime have been much investigated  (see e.g.~\cite{Follath2019,Kim2015}) and typical values in combination with Fig.~\ref{fig::inversion_spotSizeScan1} indicate that experiments may be feasible depending on the level of monochromatization. 
    Alternatively, the small focal spot can be scanned across the cavity surface from x-ray pulse to pulse. In any case,  the focusing strength plays a crucial role with regards to practical feasibility: a stronger focus is necessary for strong excitation, but simultaneously increases radiation damage. However, Fig.~\ref{fig::inversion_spotSizeScan1} shows that cavities can further help in this regard, since in free space, the fluence contours have the same slope as the excitation contours, but for the optimized cavities, the slope of the latter is changed. 
	}
	
	In summary, we have shown that optimized thin-film cavities provide means to significantly boost M\"ossbauer nuclei towards the nonlinear optics regime, in particular for conditions much below inversion and even for tight focusing. The enhancement can be interpreted  as a longitudinal compression of the pulse via the multi-path interference in the cavity. In this sense, cavities are complementary to  focusing techniques, which provide a transversal compression of the light field. We find that by combining both effects, the optimum cavity designs qualitatively differ from  state-of-the-art designs for collimated x-ray beams and optimized the cavity geometry to achieve significant enhancements down to 40~nm focusing.
	With planned upgrades to currently available sources, the resulting enhancement enables excitation of nuclei into the nonlinear regime. In particular for the isotopes $^{187}$Os and $^{169}$Tm, inversion is predicted to be achievable and $^{57}$Fe --- the workhorse of M\"ossbauer science  ---  is boosted to significantly nonlinear excitation fractions. Our results thereby outline a roadmap for further source developments and experimental efforts, balancing between advances in pulse energy and focusing capabilities.
	Progressing beyond the state-of-the-art linear regime would then allow to explore the linear quantum optical effects already observed with nuclei~\cite{Rohlsberger2010,Rohlsberger2012,Heeg2013a,Heeg2015a,Heeg2015b,Heeg2017,Haber2016a,Haber2017,Shvydko1996} in an as-yet unexplored parameter regime and potentially enable the creation of correlated x-ray photons \cite{Prasad2020,Cordier2023,Liedl2024}.
	
	\begin{acknowledgements}
		The authors would like to thank K.~P.~Heeg, O.~Diekmann, M.~Gerharz, R.~R\"ohlsberger and L.~Wolff for valuable discussions and L.~M.~Lohse for comments on an earlier version of the manuscript. DL gratefully acknowledges the Georg H. Endress Foundation for financial support and support by the DFG funded Research Training Group ``Dynamics of Controlled Atomic and Molecular Systems'' (RTG 2717).
		
		We note that this paper is based on the doctoral thesis~\cite{LentrodtPhD}. The software and figures in this paper made use of the \textsc{numpy} \cite{numpy2020}, \textsc{scipy} \cite{scipy2020} and \textsc{matplotlib} \cite{matplotlib2007} software packages.
	\end{acknowledgements}
	
	\newpage
	\appendix
	
	\section{End Matter}
\begin{figure*}[t]
		\centering
		\includegraphics[width=1.0\textwidth]{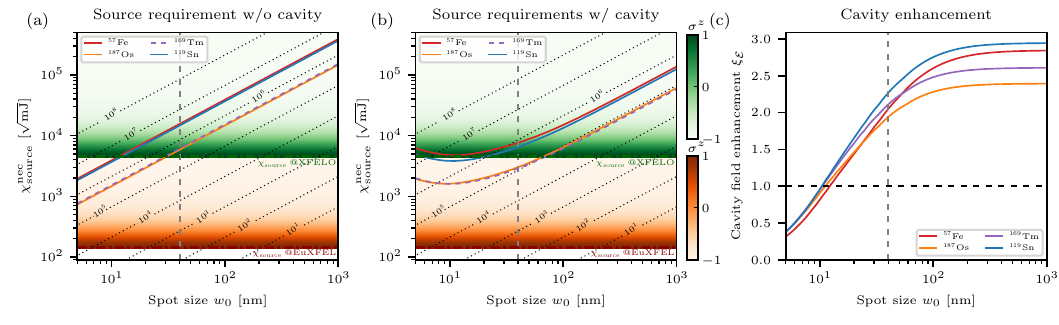}
		\caption{\newt{Source requirements for a specified cavity. This figure is analogous to Fig.~\ref{fig::inversion_spotSizeScan1}, but uses a single cavity geometry optimized for a spot size of 40$\,$nm focusing instead of at each spot size in order to illustrate the behavior of the cavity structures discussed in the main text.}}
		\label{fig::inversion_spotSizeScan2}
	\end{figure*}

	\newt{To supplement the discussion in the main text, Fig.~\ref{fig::inversion_spotSizeScan2} shows analogous results to Fig.~\ref{fig::inversion_spotSizeScan1} for a fixed cavity. That is instead of optimizing the cavity structure for each spot size, the behavior of the cavity optimized for a spot size of 40$\,$nm is investigated. In this case, the cavity enhancement levels off more rapidly, such that the necessary source parameter scales the same way as the radiation damage starting at a spot size of around 70$\,$nm. This observation further illustrates how optimized cavities can be useful to mitigate radiation damage.}

	\section{Details on the cavity optimization}
	In the optimization procedure, we use the algorithm developed in \cite{companion_nonlin2024}, which allows to compute the effect of the cavity on the pulse area of a focused x-ray pulse. The input pulse is a Gaussian beam \cite{NovotnyHecht2006} with a given beam divergence $\theta_\mathrm{div}$ and incidence angle $\theta_\mathrm{in}$.
	
	For the optimization, we fix the beam divergence such that the spot size of the beam is $40\,$nm. The cavity geometry to be optimized is assumed to have the form $m_\mathrm{clad}\,(d_1)$/C$\,$($d_2$)/$m_\mathrm{res}\,(1\,\mathrm{nm})$/C$\,$($d_3$)/$m_\mathrm{clad}\,(\infty)$. For the cladding materials, we allowed for $m_\mathrm{clad} \in$ \{Pt, Pd\}. $d_1, d_2, d_3$ were optimized as free parameters together with the incidence angle $\theta_\mathrm{in}$ and the focusing height $z_\mathrm{focus}$. That is, the geometry to be optimized is restricted to a cladding/guiding layer/substrate structure with a 1$\,$nm resonant layer inside the guiding layer. More complicated multi-layer structures are not allowed for and subject to future investigation.
	
	The optimization was then performed in \textsc{python} \cite{python} using the dual annealing global optimizer from the \textsc{scipy.optimize} \cite{scipy2020} package. Note that previous work on optimized x-ray cavity resonances includes the general study~\cite{Schenk2011PhD}. Our algorithm is designed for the particular purpose of nonlinear nuclear excitation and includes various analytical approximations that increase the efficiency.
	
	\section{Data tables}\label{app::data_tables1}
	Tables~\ref{tab::iso1}-\ref{tab::cav1} contains the data used for the inversion prediction as a reference. Each table was automatically generated using the code which was used to create Fig.~\figLettInv~in the main text.
	\begin{table*}[h]
		\begin{center} {\footnotesize
				\begin{tabular*}{\textwidth}{r|@{\extracolsep{\fill}}r|r|r|r|r|r|r|r|r|r|r|r|}
 & Z & A & E [keV] & $\gamma$ [neV] & $\alpha$ & Multipolarity & $I_g$ & $I_e$ & $d_\mathrm{eff}$ [C$\cdot$m] & $\sqrt{\sigma_\mathrm{nuc}}$ [nm] & $\sigma_\mathrm{nuc}$ [nm$^2$] & $\sigma_\mathrm{nuc}$ [kbarn]
 \\ 
 \hline 
${}^{57}$Fe & 26 & 57 & 14.41 & 4.66 & 8.6 & M1 & 0.5 & 1.5 & 3.34E-36 & 6.31E-09 & 3.99E-17 & 3.99E-10
 \\ 
${}^{193}$Pt & 78 & 193 & 1.64 & 47.00 & 3.5 & M1 & 1.5 & 0.5 & 2.01E-34 & 3.81E-07 & 1.45E-13 & 1.45E-06
 \\ 
${}^{187}$Os & 76 & 187 & 9.75 & 191.70 & 264.0 & M1 & 0.5 & 1.5 & 7.31E-36 & 1.38E-08 & 1.91E-16 & 1.91E-09
 \\ 
${}^{169}$Tm & 69 & 169 & 8.41 & 106.00 & 285.0 & M1 & 0.5 & 1.5 & 6.53E-36 & 1.23E-08 & 1.52E-16 & 1.52E-09
 \\ 
${}^{119}$Sn & 50 & 119 & 23.87 & 25.70 & 5.2 & M1 & 0.5 & 1.5 & 4.56E-36 & 8.62E-09 & 7.43E-17 & 7.43E-10
 \\ 
\end{tabular*}

			}
		\end{center}
		\caption{Isotope properties used for the inversion prediction. This table was generated automatically with the simulation code used for Fig.~\figLettInv~and is printed here as a reference. Base parameters are sourced from \cite{Rohlsberger2005,Heeg2016arxiv}.}
		\label{tab::iso1}
	\end{table*}
	\begin{table*}[h]
		\begin{center} {\footnotesize
				\begin{tabular*}{\textwidth}{r|@{\extracolsep{\fill}}r|r|r|r|r|r|r|r|}
 & Pt ($\delta$) & Pt ($\beta$) & C ($\delta$) & C ($\beta$) & res ($\delta$) & res ($\beta$) & Pd ($\delta$) & Pd ($\beta$)
 \\ 
 \hline 
${}^{57}$Fe & 1.6E-05 & 2.6E-06 & 2.3E-06 & 1.2E-09 & 7.2E-06 & 3.3E-07 & 1.0E-05 & 3.4E-07
 \\ 
${}^{193}$Pt & 8.7E-04 & 2.1E-04 & 1.7E-04 & 7.4E-06 & 1.3E-03 & 2.1E-04 & 7.0E-04 & 1.5E-04
 \\ 
${}^{187}$Os & 3.5E-05 & 2.6E-06 & 4.9E-06 & 5.4E-09 & 4.0E-05 & 2.4E-06 & 2.3E-05 & 1.4E-06
 \\ 
${}^{169}$Tm & 4.7E-05 & 4.4E-06 & 6.6E-06 & 9.9E-09 & 2.2E-05 & 1.3E-06 & 3.1E-05 & 2.5E-06
 \\ 
${}^{119}$Sn & 6.2E-06 & 4.2E-07 & 8.2E-07 & 2.8E-10 & 2.2E-06 & 3.7E-08 & 3.5E-06 & 5.2E-08
 \\ 
\end{tabular*}

			}
		\end{center}
		\caption{Refractive indices at each isotope's transition energy (see Table~\ref{tab::iso1}) used for the inversion prediction. This table was generated automatically with the simulation code used for Fig.~\figLettInv~and is printed here as a reference. Numbers are sourced from \cite{Brunetti2004,Schoonjans2011,Sturhahn2000,pynuss}.}
		\label{tab::refr}
	\end{table*}
	\begin{table*}[h]
		\begin{center} {\footnotesize
				\begin{tabular*}{\textwidth}{r|@{\extracolsep{\fill}}r|r|r|r|r|r|r|}
 & Layer 1 & Layer 2 & Layer 3 & Layer 4 & Layer 5 & $\theta_\mathrm{{in}}$ [mrad] & $z_\mathrm{{foc}}$ [nm]
 \\ 
 \hline 
${}^{57}$Fe & Pd (1.87 nm) &  C (4.37 nm) & Fe (1.00 nm) &  C (2.59 nm) & Pd ($\infty$) & 3.771 & 33.2
 \\ 
${}^{193}$Pt & Pd (2.50 nm) &  C (3.50 nm) & Pt (1.00 nm) &  C (3.50 nm) & Pd ($\infty$) & 33.110 & 0.0
 \\ 
${}^{187}$Os & Pt (1.55 nm) &  C (3.64 nm) & Os (1.00 nm) &  C (2.34 nm) & Pt ($\infty$) & 6.888 & 28.6
 \\ 
${}^{169}$Tm & Pt (1.49 nm) &  C (3.30 nm) & Tm (1.00 nm) &  C (2.44 nm) & Pt ($\infty$) & 7.494 & 28.3
 \\ 
${}^{119}$Sn & Pt (1.78 nm) &  C (2.86 nm) & Sn (1.00 nm) &  C (2.20 nm) & Pt ($\infty$) & 2.727 & 30.0
 \\ 
\end{tabular*}

			}
		\end{center}
		\caption{Optimized cavity and focusing parameters used for the inversion prediction. This table was generated automatically with the simulation code used for Fig.~\figLettInv~and is printed here as a reference.}
		\label{tab::cav1}
	\end{table*}~
	\begin{table*}[h]
		\begin{center} {\footnotesize
				\begin{tabular*}{\textwidth}{r|@{\extracolsep{\fill}}r|r|r|}
 & $E_\mathrm{pulse}$ [mJ] & $b_r$ (EuXFEL) & $b_r$ (EuXFELO, projected)
 \\ 
 \hline 
${}^{57}$Fe & 2.00E+00 & 1.00E-04 & 1.00E-07
 \\ 
${}^{193}$Pt & 2.00E+00 & 1.00E-04 & 1.00E-07
 \\ 
${}^{187}$Os & 2.00E+00 & 1.00E-04 & 1.00E-07
 \\ 
${}^{169}$Tm & 2.00E+00 & 1.00E-04 & 1.00E-07
 \\ 
${}^{119}$Sn & 2.00E+00 & 1.00E-04 & 1.00E-07
 \\ 
\end{tabular*}

			}
		\end{center}
		\caption{Source parameters used for the horizontal lines in Fig.~\figLettInv. The input parameters are $E_\mathrm{pulse}$ and $b_r$, which can be converted into the other shown parameters for a Gaussian pulse. This table was generated automatically with the simulation code used for Fig.~\figLettInv~and is printed here as a reference. Numbers are sourced from \cite{website_MID_calls}, the oscillator upgrade values are estimates from \cite{Adams2019}.}
		\label{tab::pulse}
	\end{table*}
	
	\bibliographystyle{myprsty}
	\bibliography{library}

\begin{thebibliography}{83}%
\makeatletter
\providecommand \@ifxundefined [1]{%
 \@ifx{#1\undefined}
}%
\providecommand \@ifnum [1]{%
 \ifnum #1\expandafter \@firstoftwo
 \else \expandafter \@secondoftwo
 \fi
}%
\providecommand \@ifx [1]{%
 \ifx #1\expandafter \@firstoftwo
 \else \expandafter \@secondoftwo
 \fi
}%
\providecommand \natexlab [1]{#1}%
\providecommand \enquote  [1]{``#1''}%
\providecommand \bibnamefont  [1]{#1}%
\providecommand \bibfnamefont [1]{#1}%
\providecommand \citenamefont [1]{#1}%
\providecommand \href@noop [0]{\@secondoftwo}%
\providecommand \href [0]{\begingroup \@sanitize@url \@href}%
\providecommand \@href[1]{\@@startlink{#1}\@@href}%
\providecommand \@@href[1]{\endgroup#1\@@endlink}%
\providecommand \@sanitize@url [0]{\catcode `\\12\catcode `\$12\catcode
  `\&12\catcode `\#12\catcode `\^12\catcode `\_12\catcode `\%12\relax}%
\providecommand \@@startlink[1]{}%
\providecommand \@@endlink[0]{}%
\providecommand \url  [0]{\begingroup\@sanitize@url \@url }%
\providecommand \@url [1]{\endgroup\@href {#1}{\urlprefix }}%
\providecommand \urlprefix  [0]{URL }%
\providecommand \Eprint [0]{\href }%
\providecommand \doibase [0]{http://dx.doi.org/}%
\providecommand \selectlanguage [0]{\@gobble}%
\providecommand \bibinfo  [0]{\@secondoftwo}%
\providecommand \bibfield  [0]{\@secondoftwo}%
\providecommand \translation [1]{[#1]}%
\providecommand \BibitemOpen [0]{}%
\providecommand \bibitemStop [0]{}%
\providecommand \bibitemNoStop [0]{.\EOS\space}%
\providecommand \EOS [0]{\spacefactor3000\relax}%
\providecommand \BibitemShut  [1]{\csname bibitem#1\endcsname}%
\let\auto@bib@innerbib\@empty
\bibitem [{\citenamefont {Baldwin}\ \emph {et~al.}(1981)\citenamefont
  {Baldwin}, \citenamefont {Solem},\ and\ \citenamefont
  {Gol'danskii}}]{Baldwin1981}%
  \BibitemOpen
  \bibfield  {author} {\bibinfo {author} {\bibfnamefont {G.~C.}\ \bibnamefont
  {Baldwin}}, \bibinfo {author} {\bibfnamefont {J.~C.}\ \bibnamefont {Solem}},
  \ and\ \bibinfo {author} {\bibfnamefont {V.~I.}\ \bibnamefont
  {Gol'danskii}},\ }\bibfield  {title} {\emph {\bibinfo {title} {Approaches to
  the development of gamma-ray lasers},\ }}\href {\doibase
  10.1103/RevModPhys.53.687} {\bibfield  {journal} {\bibinfo  {journal} {Rev.
  Mod. Phys.}\ }\textbf {\bibinfo {volume} {53}},\ \bibinfo {pages} {687}
  (\bibinfo {year} {1981})}\BibitemShut {NoStop}%
\bibitem [{\citenamefont {{M}{\"o}ssbauer}(1958)}]{Moessbauer1958}%
  \BibitemOpen
  \bibfield  {author} {\bibinfo {author} {\bibfnamefont {R.~L.}\ \bibnamefont
  {{M}{\"o}ssbauer}},\ }\bibfield  {title} {\emph {\bibinfo {title}
  {Kernresonanzfluoreszenz von gammastrahlung in ir191},\ }}\href {\doibase
  10.1007/BF01344210} {\bibfield  {journal} {\bibinfo  {journal} {Zeitschrift
  f{\"u}r Physik}\ }\textbf {\bibinfo {volume} {151}},\ \bibinfo {pages} {124}
  (\bibinfo {year} {1958})}\BibitemShut {NoStop}%
\bibitem [{\citenamefont {{M}{\"o}ssbauer}(1962)}]{Moessbauer1962}%
  \BibitemOpen
  \bibfield  {author} {\bibinfo {author} {\bibfnamefont {R.~L.}\ \bibnamefont
  {{M}{\"o}ssbauer}},\ }\bibfield  {title} {\emph {\bibinfo {title} {Recoilless
  nuclear resonance absorption of gamma radiation},\ }}\href {\doibase
  10.1126/science.137.3532.731} {\bibfield  {journal} {\bibinfo  {journal}
  {Science}\ }\textbf {\bibinfo {volume} {137}},\ \bibinfo {pages} {731}
  (\bibinfo {year} {1962})},\ \Eprint
  {http://arxiv.org/abs/https://science.sciencemag.org/content/137/3532/731.full.pdf}
  {https://science.sciencemag.org/content/137/3532/731.full.pdf} \BibitemShut
  {NoStop}%
\bibitem [{\citenamefont {Kocharovskaya}(1992)}]{Kocharovskaya1992}%
  \BibitemOpen
  \bibfield  {author} {\bibinfo {author} {\bibfnamefont {O.}~\bibnamefont
  {Kocharovskaya}},\ }\bibfield  {title} {\emph {\bibinfo {title}
  {Amplification and lasing without inversion},\ }}\href
  {https://www.sciencedirect.com/science/article/pii/037015739290135M}
  {\bibfield  {journal} {\bibinfo  {journal} {Physics Reports}\ }\textbf
  {\bibinfo {volume} {219}},\ \bibinfo {pages} {175} (\bibinfo {year}
  {1992})}\BibitemShut {NoStop}%
\bibitem [{\citenamefont {Baldwin}\ and\ \citenamefont
  {Solem}(1997)}]{Baldwin1997}%
  \BibitemOpen
  \bibfield  {author} {\bibinfo {author} {\bibfnamefont {G.~C.}\ \bibnamefont
  {Baldwin}}\ and\ \bibinfo {author} {\bibfnamefont {J.~C.}\ \bibnamefont
  {Solem}},\ }\bibfield  {title} {\emph {\bibinfo {title} {Recoilless gamma-ray
  lasers},\ }}\href {\doibase 10.1103/RevModPhys.69.1085} {\bibfield  {journal}
  {\bibinfo  {journal} {Rev. Mod. Phys.}\ }\textbf {\bibinfo {volume} {69}},\
  \bibinfo {pages} {1085} (\bibinfo {year} {1997})}\BibitemShut {NoStop}%
\bibitem [{\citenamefont {Pellegrini}\ \emph {et~al.}(2016)\citenamefont
  {Pellegrini}, \citenamefont {Marinelli},\ and\ \citenamefont
  {Reiche}}]{Pellegrini2016}%
  \BibitemOpen
  \bibfield  {author} {\bibinfo {author} {\bibfnamefont {C.}~\bibnamefont
  {Pellegrini}}, \bibinfo {author} {\bibfnamefont {A.}~\bibnamefont
  {Marinelli}}, \ and\ \bibinfo {author} {\bibfnamefont {S.}~\bibnamefont
  {Reiche}},\ }\bibfield  {title} {\emph {\bibinfo {title} {The physics of
  x-ray free-electron lasers},\ }}\href {\doibase 10.1103/RevModPhys.88.015006}
  {\bibfield  {journal} {\bibinfo  {journal} {Rev. Mod. Phys.}\ }\textbf
  {\bibinfo {volume} {88}},\ \bibinfo {pages} {015006} (\bibinfo {year}
  {2016})}\BibitemShut {NoStop}%
\bibitem [{\citenamefont {Georgescu}(2020)}]{Georgescu2020}%
  \BibitemOpen
  \bibfield  {author} {\bibinfo {author} {\bibfnamefont {I.}~\bibnamefont
  {Georgescu}},\ }\bibfield  {title} {\emph {\bibinfo {title} {The first decade
  of xfels},\ }}\href {\doibase 10.1038/s42254-020-0204-6} {\bibfield
  {journal} {\bibinfo  {journal} {Nature Reviews Physics}\ }\textbf {\bibinfo
  {volume} {2}},\ \bibinfo {pages} {345} (\bibinfo {year} {2020})}\BibitemShut
  {NoStop}%
\bibitem [{\citenamefont {Schawlow}\ and\ \citenamefont
  {Townes}(1958)}]{Schawlow1958}%
  \BibitemOpen
  \bibfield  {author} {\bibinfo {author} {\bibfnamefont {A.~L.}\ \bibnamefont
  {Schawlow}}\ and\ \bibinfo {author} {\bibfnamefont {C.~H.}\ \bibnamefont
  {Townes}},\ }\bibfield  {title} {\emph {\bibinfo {title} {Infrared and
  optical masers},\ }}\href {\doibase 10.1103/PhysRev.112.1940} {\bibfield
  {journal} {\bibinfo  {journal} {Phys. Rev.}\ }\textbf {\bibinfo {volume}
  {112}},\ \bibinfo {pages} {1940} (\bibinfo {year} {1958})}\BibitemShut
  {NoStop}%
\bibitem [{\citenamefont {Lamb}\ \emph {et~al.}(1999)\citenamefont {Lamb},
  \citenamefont {Schleich}, \citenamefont {Scully},\ and\ \citenamefont
  {Townes}}]{Lamb1999}%
  \BibitemOpen
  \bibfield  {author} {\bibinfo {author} {\bibfnamefont {W.~E.}\ \bibnamefont
  {Lamb}}, \bibinfo {author} {\bibfnamefont {W.~P.}\ \bibnamefont {Schleich}},
  \bibinfo {author} {\bibfnamefont {M.~O.}\ \bibnamefont {Scully}}, \ and\
  \bibinfo {author} {\bibfnamefont {C.~H.}\ \bibnamefont {Townes}},\ }\bibfield
   {title} {\emph {\bibinfo {title} {Laser physics: Quantum controversy in
  action},\ }}\href {\doibase 10.1103/RevModPhys.71.S263} {\bibfield  {journal}
  {\bibinfo  {journal} {Rev. Mod. Phys.}\ }\textbf {\bibinfo {volume} {71}},\
  \bibinfo {pages} {S263} (\bibinfo {year} {1999})}\BibitemShut {NoStop}%
\bibitem [{\citenamefont {Matsuda}\ and\ \citenamefont
  {Arafune}(2023)}]{Matsuda2023}%
  \BibitemOpen
  \bibinfo {editor} {\bibfnamefont {I.}~\bibnamefont {Matsuda}}\ and\ \bibinfo
  {editor} {\bibfnamefont {R.}~\bibnamefont {Arafune}},\ eds.,\ \href {\doibase
  10.1007/978-981-99-6714-8} {\emph {\bibinfo {title} {{N}onlinear {X}-ray
  spectroscopy for materials science}}},\ \bibinfo {series} {Springer series in
  optical sciences}, Vol.\ \bibinfo {volume} {246}\ (\bibinfo  {publisher}
  {Springer},\ \bibinfo {address} {Singapore},\ \bibinfo {year} {2023})\ pp.\
  \bibinfo {pages} {xiv, 160 pages : illustrations, diagrams}\BibitemShut
  {NoStop}%
\bibitem [{\citenamefont {Rohringer}(2019)}]{Rohringer2019}%
  \BibitemOpen
  \bibfield  {author} {\bibinfo {author} {\bibfnamefont {N.}~\bibnamefont
  {Rohringer}},\ }\bibfield  {title} {\emph {\bibinfo {title} {X-ray raman
  scattering: a building block for nonlinear spectroscopy},\ }}\href {\doibase
  10.1098/rsta.2017.0471} {\bibfield  {journal} {\bibinfo  {journal}
  {Philosophical Transactions of the Royal Society A: Mathematical, Physical
  and Engineering Sciences}\ }\textbf {\bibinfo {volume} {377}},\ \bibinfo
  {pages} {20170471} (\bibinfo {year} {2019})}\BibitemShut {NoStop}%
\bibitem [{\citenamefont {Sofer}\ \emph {et~al.}(2019)\citenamefont {Sofer},
  \citenamefont {Strizhevsky}, \citenamefont {Schori}, \citenamefont
  {Tamasaku},\ and\ \citenamefont {Shwartz}}]{Sofer2019}%
  \BibitemOpen
  \bibfield  {author} {\bibinfo {author} {\bibfnamefont {S.}~\bibnamefont
  {Sofer}}, \bibinfo {author} {\bibfnamefont {E.}~\bibnamefont {Strizhevsky}},
  \bibinfo {author} {\bibfnamefont {A.}~\bibnamefont {Schori}}, \bibinfo
  {author} {\bibfnamefont {K.}~\bibnamefont {Tamasaku}}, \ and\ \bibinfo
  {author} {\bibfnamefont {S.}~\bibnamefont {Shwartz}},\ }\bibfield  {title}
  {\emph {\bibinfo {title} {Quantum enhanced x-ray detection},\ }}\href
  {\doibase 10.1103/PhysRevX.9.031033} {\bibfield  {journal} {\bibinfo
  {journal} {Phys. Rev. X}\ }\textbf {\bibinfo {volume} {9}},\ \bibinfo {pages}
  {031033} (\bibinfo {year} {2019})}\BibitemShut {NoStop}%
\bibitem [{\citenamefont {Nandi}\ \emph {et~al.}(2022)\citenamefont {Nandi},
  \citenamefont {Olofsson}, \citenamefont {Bertolino}, \citenamefont
  {Carlstr{\"o}m}, \citenamefont {Zapata}, \citenamefont {Busto}, \citenamefont
  {Callegari}, \citenamefont {Di~Fraia}, \citenamefont {Eng-Johnsson},
  \citenamefont {Feifel}, \citenamefont {Gallician}, \citenamefont
  {Gisselbrecht}, \citenamefont {Maclot}, \citenamefont {Neori{\v{c}}i{\'{c}}},
  \citenamefont {Peschel}, \citenamefont {Plekan}, \citenamefont {Prince},
  \citenamefont {Squibb}, \citenamefont {Zhong}, \citenamefont {Demekhin},
  \citenamefont {Meyer}, \citenamefont {Miron}, \citenamefont {Badano},
  \citenamefont {Danailov}, \citenamefont {Giannessi}, \citenamefont
  {Manfredda}, \citenamefont {Sottocorona}, \citenamefont {Zangrando},\ and\
  \citenamefont {Dahlstr{\"o}m}}]{Nandi2022}%
  \BibitemOpen
  \bibfield  {author} {\bibinfo {author} {\bibfnamefont {S.}~\bibnamefont
  {Nandi}}, \bibinfo {author} {\bibfnamefont {E.}~\bibnamefont {Olofsson}},
  \bibinfo {author} {\bibfnamefont {M.}~\bibnamefont {Bertolino}}, \bibinfo
  {author} {\bibfnamefont {S.}~\bibnamefont {Carlstr{\"o}m}}, \bibinfo {author}
  {\bibfnamefont {F.}~\bibnamefont {Zapata}}, \bibinfo {author} {\bibfnamefont
  {D.}~\bibnamefont {Busto}}, \bibinfo {author} {\bibfnamefont
  {C.}~\bibnamefont {Callegari}}, \bibinfo {author} {\bibfnamefont
  {M.}~\bibnamefont {Di~Fraia}}, \bibinfo {author} {\bibfnamefont
  {P.}~\bibnamefont {Eng-Johnsson}}, \bibinfo {author} {\bibfnamefont
  {R.}~\bibnamefont {Feifel}}, \bibinfo {author} {\bibfnamefont
  {G.}~\bibnamefont {Gallician}}, \bibinfo {author} {\bibfnamefont
  {M.}~\bibnamefont {Gisselbrecht}}, \bibinfo {author} {\bibfnamefont
  {S.}~\bibnamefont {Maclot}}, \bibinfo {author} {\bibfnamefont
  {L.}~\bibnamefont {Neori{\v{c}}i{\'{c}}}}, \bibinfo {author} {\bibfnamefont
  {J.}~\bibnamefont {Peschel}}, \bibinfo {author} {\bibfnamefont
  {O.}~\bibnamefont {Plekan}}, \bibinfo {author} {\bibfnamefont {K.~C.}\
  \bibnamefont {Prince}}, \bibinfo {author} {\bibfnamefont {R.~J.}\
  \bibnamefont {Squibb}}, \bibinfo {author} {\bibfnamefont {S.}~\bibnamefont
  {Zhong}}, \bibinfo {author} {\bibfnamefont {P.~V.}\ \bibnamefont {Demekhin}},
  \bibinfo {author} {\bibfnamefont {M.}~\bibnamefont {Meyer}}, \bibinfo
  {author} {\bibfnamefont {C.}~\bibnamefont {Miron}}, \bibinfo {author}
  {\bibfnamefont {L.}~\bibnamefont {Badano}}, \bibinfo {author} {\bibfnamefont
  {M.~B.}\ \bibnamefont {Danailov}}, \bibinfo {author} {\bibfnamefont
  {L.}~\bibnamefont {Giannessi}}, \bibinfo {author} {\bibfnamefont
  {M.}~\bibnamefont {Manfredda}}, \bibinfo {author} {\bibfnamefont
  {F.}~\bibnamefont {Sottocorona}}, \bibinfo {author} {\bibfnamefont
  {M.}~\bibnamefont {Zangrando}}, \ and\ \bibinfo {author} {\bibfnamefont
  {J.~M.}\ \bibnamefont {Dahlstr{\"o}m}},\ }\bibfield  {title} {\emph {\bibinfo
  {title} {Observation of rabi dynamics with a short-wavelength free-electron
  laser},\ }}\href {\doibase 10.1038/s41586-022-04948-y} {\bibfield  {journal}
  {\bibinfo  {journal} {Nature}\ }\textbf {\bibinfo {volume} {608}},\ \bibinfo
  {pages} {488} (\bibinfo {year} {2022})}\BibitemShut {NoStop}%
\bibitem [{\citenamefont {Doumy}\ \emph {et~al.}(2011)\citenamefont {Doumy},
  \citenamefont {Roedig}, \citenamefont {Son}, \citenamefont {Blaga},
  \citenamefont {DiChiara}, \citenamefont {Santra}, \citenamefont {Berrah},
  \citenamefont {Bostedt}, \citenamefont {Bozek}, \citenamefont {Bucksbaum},
  \citenamefont {Cryan}, \citenamefont {Fang}, \citenamefont {Ghimire},
  \citenamefont {Glownia}, \citenamefont {Hoener}, \citenamefont {Kanter},
  \citenamefont {Kr\"assig}, \citenamefont {Kuebel}, \citenamefont
  {Messerschmidt}, \citenamefont {Paulus}, \citenamefont {Reis}, \citenamefont
  {Rohringer}, \citenamefont {Young}, \citenamefont {Agostini},\ and\
  \citenamefont {DiMauro}}]{Doumy2011}%
  \BibitemOpen
  \bibfield  {author} {\bibinfo {author} {\bibfnamefont {G.}~\bibnamefont
  {Doumy}}, \bibinfo {author} {\bibfnamefont {C.}~\bibnamefont {Roedig}},
  \bibinfo {author} {\bibfnamefont {S.-K.}\ \bibnamefont {Son}}, \bibinfo
  {author} {\bibfnamefont {C.~I.}\ \bibnamefont {Blaga}}, \bibinfo {author}
  {\bibfnamefont {A.~D.}\ \bibnamefont {DiChiara}}, \bibinfo {author}
  {\bibfnamefont {R.}~\bibnamefont {Santra}}, \bibinfo {author} {\bibfnamefont
  {N.}~\bibnamefont {Berrah}}, \bibinfo {author} {\bibfnamefont
  {C.}~\bibnamefont {Bostedt}}, \bibinfo {author} {\bibfnamefont {J.~D.}\
  \bibnamefont {Bozek}}, \bibinfo {author} {\bibfnamefont {P.~H.}\ \bibnamefont
  {Bucksbaum}}, \bibinfo {author} {\bibfnamefont {J.~P.}\ \bibnamefont
  {Cryan}}, \bibinfo {author} {\bibfnamefont {L.}~\bibnamefont {Fang}},
  \bibinfo {author} {\bibfnamefont {S.}~\bibnamefont {Ghimire}}, \bibinfo
  {author} {\bibfnamefont {J.~M.}\ \bibnamefont {Glownia}}, \bibinfo {author}
  {\bibfnamefont {M.}~\bibnamefont {Hoener}}, \bibinfo {author} {\bibfnamefont
  {E.~P.}\ \bibnamefont {Kanter}}, \bibinfo {author} {\bibfnamefont
  {B.}~\bibnamefont {Kr\"assig}}, \bibinfo {author} {\bibfnamefont
  {M.}~\bibnamefont {Kuebel}}, \bibinfo {author} {\bibfnamefont
  {M.}~\bibnamefont {Messerschmidt}}, \bibinfo {author} {\bibfnamefont {G.~G.}\
  \bibnamefont {Paulus}}, \bibinfo {author} {\bibfnamefont {D.~A.}\
  \bibnamefont {Reis}}, \bibinfo {author} {\bibfnamefont {N.}~\bibnamefont
  {Rohringer}}, \bibinfo {author} {\bibfnamefont {L.}~\bibnamefont {Young}},
  \bibinfo {author} {\bibfnamefont {P.}~\bibnamefont {Agostini}}, \ and\
  \bibinfo {author} {\bibfnamefont {L.~F.}\ \bibnamefont {DiMauro}},\
  }\bibfield  {title} {\emph {\bibinfo {title} {Nonlinear atomic response to
  intense ultrashort x rays},\ }}\href {\doibase
  10.1103/PhysRevLett.106.083002} {\bibfield  {journal} {\bibinfo  {journal}
  {Phys. Rev. Lett.}\ }\textbf {\bibinfo {volume} {106}},\ \bibinfo {pages}
  {083002} (\bibinfo {year} {2011})}\BibitemShut {NoStop}%
\bibitem [{\citenamefont {Rohringer}\ \emph {et~al.}(2012)\citenamefont
  {Rohringer}, \citenamefont {Ryan}, \citenamefont {London}, \citenamefont
  {Purvis}, \citenamefont {Albert}, \citenamefont {Dunn}, \citenamefont
  {Bozek}, \citenamefont {Bostedt}, \citenamefont {Graf}, \citenamefont {Hill},
  \citenamefont {Hau-Riege},\ and\ \citenamefont {Rocca}}]{Rohringer2012}%
  \BibitemOpen
  \bibfield  {author} {\bibinfo {author} {\bibfnamefont {N.}~\bibnamefont
  {Rohringer}}, \bibinfo {author} {\bibfnamefont {D.}~\bibnamefont {Ryan}},
  \bibinfo {author} {\bibfnamefont {R.~A.}\ \bibnamefont {London}}, \bibinfo
  {author} {\bibfnamefont {M.}~\bibnamefont {Purvis}}, \bibinfo {author}
  {\bibfnamefont {F.}~\bibnamefont {Albert}}, \bibinfo {author} {\bibfnamefont
  {J.}~\bibnamefont {Dunn}}, \bibinfo {author} {\bibfnamefont {J.~D.}\
  \bibnamefont {Bozek}}, \bibinfo {author} {\bibfnamefont {C.}~\bibnamefont
  {Bostedt}}, \bibinfo {author} {\bibfnamefont {A.}~\bibnamefont {Graf}},
  \bibinfo {author} {\bibfnamefont {R.}~\bibnamefont {Hill}}, \bibinfo {author}
  {\bibfnamefont {S.~P.}\ \bibnamefont {Hau-Riege}}, \ and\ \bibinfo {author}
  {\bibfnamefont {J.~J.}\ \bibnamefont {Rocca}},\ }\bibfield  {title} {\emph
  {\bibinfo {title} {Atomic inner-shell x-ray laser at 1.46 nanometres pumped
  by an x-ray free-electron laser},\ }}\href {\doibase 10.1038/nature10721}
  {\bibfield  {journal} {\bibinfo  {journal} {Nature}\ }\textbf {\bibinfo
  {volume} {481}},\ \bibinfo {pages} {488} (\bibinfo {year}
  {2012})}\BibitemShut {NoStop}%
\bibitem [{\citenamefont {R{\"o}hlsberger}\ \emph {et~al.}(2010)\citenamefont
  {R{\"o}hlsberger}, \citenamefont {Schlage}, \citenamefont {Sahoo},
  \citenamefont {Couet},\ and\ \citenamefont {R{\"u}ffer}}]{Rohlsberger2010}%
  \BibitemOpen
  \bibfield  {author} {\bibinfo {author} {\bibfnamefont {R.}~\bibnamefont
  {R{\"o}hlsberger}}, \bibinfo {author} {\bibfnamefont {K.}~\bibnamefont
  {Schlage}}, \bibinfo {author} {\bibfnamefont {B.}~\bibnamefont {Sahoo}},
  \bibinfo {author} {\bibfnamefont {S.}~\bibnamefont {Couet}}, \ and\ \bibinfo
  {author} {\bibfnamefont {R.}~\bibnamefont {R{\"u}ffer}},\ }\bibfield  {title}
  {\emph {\bibinfo {title} {Collective lamb shift in single-photon
  superradiance},\ }}\href {\doibase 10.1126/science.1187770} {\bibfield
  {journal} {\bibinfo  {journal} {Science}\ }\textbf {\bibinfo {volume}
  {328}},\ \bibinfo {pages} {1248} (\bibinfo {year} {2010})}\BibitemShut
  {NoStop}%
\bibitem [{\citenamefont {R{\"{o}}hlsberger}\ \emph {et~al.}(2012)\citenamefont
  {R{\"{o}}hlsberger}, \citenamefont {Wille}, \citenamefont {Schlage},\ and\
  \citenamefont {Sahoo}}]{Rohlsberger2012}%
  \BibitemOpen
  \bibfield  {author} {\bibinfo {author} {\bibfnamefont {R.}~\bibnamefont
  {R{\"{o}}hlsberger}}, \bibinfo {author} {\bibfnamefont {H.-C.}\ \bibnamefont
  {Wille}}, \bibinfo {author} {\bibfnamefont {K.}~\bibnamefont {Schlage}}, \
  and\ \bibinfo {author} {\bibfnamefont {B.}~\bibnamefont {Sahoo}},\ }\bibfield
   {title} {\emph {\bibinfo {title} {{Electromagnetically induced transparency
  with resonant nuclei in a cavity}},\ }}\href {\doibase 10.1038/nature10741}
  {\bibfield  {journal} {\bibinfo  {journal} {Nature}\ }\textbf {\bibinfo
  {volume} {482}},\ \bibinfo {pages} {199} (\bibinfo {year}
  {2012})}\BibitemShut {NoStop}%
\bibitem [{\citenamefont {Heeg}\ \emph {et~al.}(2013)\citenamefont {Heeg},
  \citenamefont {Wille}, \citenamefont {Schlage}, \citenamefont {Guryeva},
  \citenamefont {Schumacher}, \citenamefont {Uschmann}, \citenamefont
  {Schulze}, \citenamefont {Marx}, \citenamefont {K\"ampfer}, \citenamefont
  {Paulus}, \citenamefont {R\"ohlsberger},\ and\ \citenamefont
  {Evers}}]{Heeg2013a}%
  \BibitemOpen
  \bibfield  {author} {\bibinfo {author} {\bibfnamefont {K.~P.}\ \bibnamefont
  {Heeg}}, \bibinfo {author} {\bibfnamefont {H.-C.}\ \bibnamefont {Wille}},
  \bibinfo {author} {\bibfnamefont {K.}~\bibnamefont {Schlage}}, \bibinfo
  {author} {\bibfnamefont {T.}~\bibnamefont {Guryeva}}, \bibinfo {author}
  {\bibfnamefont {D.}~\bibnamefont {Schumacher}}, \bibinfo {author}
  {\bibfnamefont {I.}~\bibnamefont {Uschmann}}, \bibinfo {author}
  {\bibfnamefont {K.~S.}\ \bibnamefont {Schulze}}, \bibinfo {author}
  {\bibfnamefont {B.}~\bibnamefont {Marx}}, \bibinfo {author} {\bibfnamefont
  {T.}~\bibnamefont {K\"ampfer}}, \bibinfo {author} {\bibfnamefont {G.~G.}\
  \bibnamefont {Paulus}}, \bibinfo {author} {\bibfnamefont {R.}~\bibnamefont
  {R\"ohlsberger}}, \ and\ \bibinfo {author} {\bibfnamefont {J.}~\bibnamefont
  {Evers}},\ }\bibfield  {title} {\emph {\bibinfo {title} {Vacuum-assisted
  generation and control of atomic coherences at x-ray energies},\ }}\href
  {\doibase 10.1103/PhysRevLett.111.073601} {\bibfield  {journal} {\bibinfo
  {journal} {Phys. Rev. Lett.}\ }\textbf {\bibinfo {volume} {111}},\ \bibinfo
  {pages} {073601} (\bibinfo {year} {2013})}\BibitemShut {NoStop}%
\bibitem [{\citenamefont {Heeg}\ \emph
  {et~al.}(2015{\natexlab{a}})\citenamefont {Heeg}, \citenamefont {Ott},
  \citenamefont {Schumacher}, \citenamefont {Wille}, \citenamefont
  {R\"ohlsberger}, \citenamefont {Pfeifer},\ and\ \citenamefont
  {Evers}}]{Heeg2015a}%
  \BibitemOpen
  \bibfield  {author} {\bibinfo {author} {\bibfnamefont {K.~P.}\ \bibnamefont
  {Heeg}}, \bibinfo {author} {\bibfnamefont {C.}~\bibnamefont {Ott}}, \bibinfo
  {author} {\bibfnamefont {D.}~\bibnamefont {Schumacher}}, \bibinfo {author}
  {\bibfnamefont {H.-C.}\ \bibnamefont {Wille}}, \bibinfo {author}
  {\bibfnamefont {R.}~\bibnamefont {R\"ohlsberger}}, \bibinfo {author}
  {\bibfnamefont {T.}~\bibnamefont {Pfeifer}}, \ and\ \bibinfo {author}
  {\bibfnamefont {J.}~\bibnamefont {Evers}},\ }\bibfield  {title} {\emph
  {\bibinfo {title} {Interferometric phase detection at x-ray energies via
  {F}ano resonance control},\ }}\href {\doibase 10.1103/PhysRevLett.114.207401}
  {\bibfield  {journal} {\bibinfo  {journal} {Phys. Rev. Lett.}\ }\textbf
  {\bibinfo {volume} {114}},\ \bibinfo {pages} {207401} (\bibinfo {year}
  {2015}{\natexlab{a}})}\BibitemShut {NoStop}%
\bibitem [{\citenamefont {Heeg}\ \emph
  {et~al.}(2015{\natexlab{b}})\citenamefont {Heeg}, \citenamefont {Haber},
  \citenamefont {Schumacher}, \citenamefont {Bocklage}, \citenamefont {Wille},
  \citenamefont {Schulze}, \citenamefont {Loetzsch}, \citenamefont {Uschmann},
  \citenamefont {Paulus}, \citenamefont {R\"uffer}, \citenamefont
  {R\"ohlsberger},\ and\ \citenamefont {Evers}}]{Heeg2015b}%
  \BibitemOpen
  \bibfield  {author} {\bibinfo {author} {\bibfnamefont {K.~P.}\ \bibnamefont
  {Heeg}}, \bibinfo {author} {\bibfnamefont {J.}~\bibnamefont {Haber}},
  \bibinfo {author} {\bibfnamefont {D.}~\bibnamefont {Schumacher}}, \bibinfo
  {author} {\bibfnamefont {L.}~\bibnamefont {Bocklage}}, \bibinfo {author}
  {\bibfnamefont {H.-C.}\ \bibnamefont {Wille}}, \bibinfo {author}
  {\bibfnamefont {K.~S.}\ \bibnamefont {Schulze}}, \bibinfo {author}
  {\bibfnamefont {R.}~\bibnamefont {Loetzsch}}, \bibinfo {author}
  {\bibfnamefont {I.}~\bibnamefont {Uschmann}}, \bibinfo {author}
  {\bibfnamefont {G.~G.}\ \bibnamefont {Paulus}}, \bibinfo {author}
  {\bibfnamefont {R.}~\bibnamefont {R\"uffer}}, \bibinfo {author}
  {\bibfnamefont {R.}~\bibnamefont {R\"ohlsberger}}, \ and\ \bibinfo {author}
  {\bibfnamefont {J.}~\bibnamefont {Evers}},\ }\bibfield  {title} {\emph
  {\bibinfo {title} {Tunable subluminal propagation of narrow-band x-ray
  pulses},\ }}\href {\doibase 10.1103/PhysRevLett.114.203601} {\bibfield
  {journal} {\bibinfo  {journal} {Phys. Rev. Lett.}\ }\textbf {\bibinfo
  {volume} {114}},\ \bibinfo {pages} {203601} (\bibinfo {year}
  {2015}{\natexlab{b}})}\BibitemShut {NoStop}%
\bibitem [{\citenamefont {Haber}\ \emph {et~al.}(2016)\citenamefont {Haber},
  \citenamefont {Schulze}, \citenamefont {Schlage}, \citenamefont {Loetzsch},
  \citenamefont {Bocklage}, \citenamefont {Gurieva}, \citenamefont {Bernhardt},
  \citenamefont {Wille}, \citenamefont {R{\"u}ffer}, \citenamefont {Uschmann},
  \citenamefont {Paulus},\ and\ \citenamefont {R{\"o}hlsberger}}]{Haber2016a}%
  \BibitemOpen
  \bibfield  {author} {\bibinfo {author} {\bibfnamefont {J.}~\bibnamefont
  {Haber}}, \bibinfo {author} {\bibfnamefont {K.~S.}\ \bibnamefont {Schulze}},
  \bibinfo {author} {\bibfnamefont {K.}~\bibnamefont {Schlage}}, \bibinfo
  {author} {\bibfnamefont {R.}~\bibnamefont {Loetzsch}}, \bibinfo {author}
  {\bibfnamefont {L.}~\bibnamefont {Bocklage}}, \bibinfo {author}
  {\bibfnamefont {T.}~\bibnamefont {Gurieva}}, \bibinfo {author} {\bibfnamefont
  {H.}~\bibnamefont {Bernhardt}}, \bibinfo {author} {\bibfnamefont {H.-C.}\
  \bibnamefont {Wille}}, \bibinfo {author} {\bibfnamefont {R.}~\bibnamefont
  {R{\"u}ffer}}, \bibinfo {author} {\bibfnamefont {I.}~\bibnamefont
  {Uschmann}}, \bibinfo {author} {\bibfnamefont {G.~G.}\ \bibnamefont
  {Paulus}}, \ and\ \bibinfo {author} {\bibfnamefont {R.}~\bibnamefont
  {R{\"o}hlsberger}},\ }\bibfield  {title} {\emph {\bibinfo {title} {Collective
  strong coupling of x-rays and nuclei in a nuclear optical lattice},\ }}\href
  {https://doi.org/10.1038/nphoton.2016.77} {\bibfield  {journal} {\bibinfo
  {journal} {Nat. Phot.}\ }\textbf {\bibinfo {volume} {10}},\ \bibinfo {pages}
  {445 EP } (\bibinfo {year} {2016})}\BibitemShut {NoStop}%
\bibitem [{\citenamefont {Haber}\ \emph {et~al.}(2017)\citenamefont {Haber},
  \citenamefont {Kong}, \citenamefont {Strohm}, \citenamefont {Willing},
  \citenamefont {Gollwitzer}, \citenamefont {Bocklage}, \citenamefont
  {R{\"u}ffer}, \citenamefont {P{\'a}lffy},\ and\ \citenamefont
  {R{\"o}hlsberger}}]{Haber2017}%
  \BibitemOpen
  \bibfield  {author} {\bibinfo {author} {\bibfnamefont {J.}~\bibnamefont
  {Haber}}, \bibinfo {author} {\bibfnamefont {X.}~\bibnamefont {Kong}},
  \bibinfo {author} {\bibfnamefont {C.}~\bibnamefont {Strohm}}, \bibinfo
  {author} {\bibfnamefont {S.}~\bibnamefont {Willing}}, \bibinfo {author}
  {\bibfnamefont {J.}~\bibnamefont {Gollwitzer}}, \bibinfo {author}
  {\bibfnamefont {L.}~\bibnamefont {Bocklage}}, \bibinfo {author}
  {\bibfnamefont {R.}~\bibnamefont {R{\"u}ffer}}, \bibinfo {author}
  {\bibfnamefont {A.}~\bibnamefont {P{\'a}lffy}}, \ and\ \bibinfo {author}
  {\bibfnamefont {R.}~\bibnamefont {R{\"o}hlsberger}},\ }\bibfield  {title}
  {\emph {\bibinfo {title} {{R}abi oscillations of x-ray radiation between two
  nuclear ensembles},\ }}\href {\doibase 10.1038/s41566-017-0013-3} {\bibfield
  {journal} {\bibinfo  {journal} {Nat. Phot.}\ }\textbf {\bibinfo {volume}
  {11}},\ \bibinfo {pages} {720} (\bibinfo {year} {2017})}\BibitemShut
  {NoStop}%
\bibitem [{\citenamefont {Helist{\"o}}\ \emph {et~al.}(1991)\citenamefont
  {Helist{\"o}}, \citenamefont {Tittonen}, \citenamefont {Lippmaa},\ and\
  \citenamefont {Katila}}]{Helisto1991}%
  \BibitemOpen
  \bibfield  {author} {\bibinfo {author} {\bibfnamefont {P.}~\bibnamefont
  {Helist{\"o}}}, \bibinfo {author} {\bibfnamefont {I.}~\bibnamefont
  {Tittonen}}, \bibinfo {author} {\bibfnamefont {M.}~\bibnamefont {Lippmaa}}, \
  and\ \bibinfo {author} {\bibfnamefont {T.}~\bibnamefont {Katila}},\
  }\bibfield  {title} {\emph {\bibinfo {title} {Gamma echo},\ }}\href {\doibase
  10.1103/PhysRevLett.66.2037} {\bibfield  {journal} {\bibinfo  {journal}
  {Phys. Rev. Lett.}\ }\textbf {\bibinfo {volume} {66}},\ \bibinfo {pages}
  {2037} (\bibinfo {year} {1991})}\BibitemShut {NoStop}%
\bibitem [{\citenamefont {Shvyd'ko}\ \emph {et~al.}(1996)\citenamefont
  {Shvyd'ko}, \citenamefont {Hertrich}, \citenamefont {van B\"urck},
  \citenamefont {Gerdau}, \citenamefont {Leupold}, \citenamefont {Metge},
  \citenamefont {R\"uter}, \citenamefont {Schwendy}, \citenamefont {Smirnov},
  \citenamefont {Potzel},\ and\ \citenamefont {Schindelmann}}]{Shvydko1996}%
  \BibitemOpen
  \bibfield  {author} {\bibinfo {author} {\bibfnamefont {Y.~V.}\ \bibnamefont
  {Shvyd'ko}}, \bibinfo {author} {\bibfnamefont {T.}~\bibnamefont {Hertrich}},
  \bibinfo {author} {\bibfnamefont {U.}~\bibnamefont {van B\"urck}}, \bibinfo
  {author} {\bibfnamefont {E.}~\bibnamefont {Gerdau}}, \bibinfo {author}
  {\bibfnamefont {O.}~\bibnamefont {Leupold}}, \bibinfo {author} {\bibfnamefont
  {J.}~\bibnamefont {Metge}}, \bibinfo {author} {\bibfnamefont {H.~D.}\
  \bibnamefont {R\"uter}}, \bibinfo {author} {\bibfnamefont {S.}~\bibnamefont
  {Schwendy}}, \bibinfo {author} {\bibfnamefont {G.~V.}\ \bibnamefont
  {Smirnov}}, \bibinfo {author} {\bibfnamefont {W.}~\bibnamefont {Potzel}}, \
  and\ \bibinfo {author} {\bibfnamefont {P.}~\bibnamefont {Schindelmann}},\
  }\bibfield  {title} {\emph {\bibinfo {title} {Storage of nuclear excitation
  energy through magnetic switching},\ }}\href {\doibase
  10.1103/PhysRevLett.77.3232} {\bibfield  {journal} {\bibinfo  {journal}
  {Phys. Rev. Lett.}\ }\textbf {\bibinfo {volume} {77}},\ \bibinfo {pages}
  {3232} (\bibinfo {year} {1996})}\BibitemShut {NoStop}%
\bibitem [{\citenamefont {Schindelmann}\ \emph {et~al.}(2002)\citenamefont
  {Schindelmann}, \citenamefont {van B\"urck}, \citenamefont {Potzel},
  \citenamefont {Smirnov}, \citenamefont {Popov}, \citenamefont {Gerdau},
  \citenamefont {Shvyd'ko}, \citenamefont {J\"aschke}, \citenamefont {R\"uter},
  \citenamefont {Chumakov},\ and\ \citenamefont {R\"uffer}}]{Schindelmann2002}%
  \BibitemOpen
  \bibfield  {author} {\bibinfo {author} {\bibfnamefont {P.}~\bibnamefont
  {Schindelmann}}, \bibinfo {author} {\bibfnamefont {U.}~\bibnamefont {van
  B\"urck}}, \bibinfo {author} {\bibfnamefont {W.}~\bibnamefont {Potzel}},
  \bibinfo {author} {\bibfnamefont {G.~V.}\ \bibnamefont {Smirnov}}, \bibinfo
  {author} {\bibfnamefont {S.~L.}\ \bibnamefont {Popov}}, \bibinfo {author}
  {\bibfnamefont {E.}~\bibnamefont {Gerdau}}, \bibinfo {author} {\bibfnamefont
  {Y.~V.}\ \bibnamefont {Shvyd'ko}}, \bibinfo {author} {\bibfnamefont
  {J.}~\bibnamefont {J\"aschke}}, \bibinfo {author} {\bibfnamefont {H.~D.}\
  \bibnamefont {R\"uter}}, \bibinfo {author} {\bibfnamefont {A.~I.}\
  \bibnamefont {Chumakov}}, \ and\ \bibinfo {author} {\bibfnamefont
  {R.}~\bibnamefont {R\"uffer}},\ }\bibfield  {title} {\emph {\bibinfo {title}
  {Radiative decoupling and coupling of nuclear oscillators by stepwise
  doppler-energy shifts},\ }}\href {\doibase 10.1103/PhysRevA.65.023804}
  {\bibfield  {journal} {\bibinfo  {journal} {Phys. Rev. A}\ }\textbf {\bibinfo
  {volume} {65}},\ \bibinfo {pages} {023804} (\bibinfo {year}
  {2002})}\BibitemShut {NoStop}%
\bibitem [{\citenamefont {Vagizov}\ \emph {et~al.}(2014)\citenamefont
  {Vagizov}, \citenamefont {Antonov}, \citenamefont {Radeonychev},
  \citenamefont {Shakhmuratov},\ and\ \citenamefont
  {Kocharovskaya}}]{Vagizov2014}%
  \BibitemOpen
  \bibfield  {author} {\bibinfo {author} {\bibfnamefont {F.}~\bibnamefont
  {Vagizov}}, \bibinfo {author} {\bibfnamefont {V.}~\bibnamefont {Antonov}},
  \bibinfo {author} {\bibfnamefont {Y.~V.}\ \bibnamefont {Radeonychev}},
  \bibinfo {author} {\bibfnamefont {R.~N.}\ \bibnamefont {Shakhmuratov}}, \
  and\ \bibinfo {author} {\bibfnamefont {O.}~\bibnamefont {Kocharovskaya}},\
  }\bibfield  {title} {\emph {\bibinfo {title} {Coherent control of the
  waveforms of recoilless g-ray photons},\ }}\href {\doibase
  10.1038/nature13018} {\bibfield  {journal} {\bibinfo  {journal} {Nature}\
  }\textbf {\bibinfo {volume} {508}},\ \bibinfo {pages} {80} (\bibinfo {year}
  {2014})}\BibitemShut {NoStop}%
\bibitem [{\citenamefont {Sakshath}\ \emph {et~al.}(2017)\citenamefont
  {Sakshath}, \citenamefont {Jenni}, \citenamefont {Scherthan}, \citenamefont
  {W{\"u}rtz}, \citenamefont {Herlitschke}, \citenamefont {Sergeev},
  \citenamefont {Strohm}, \citenamefont {Wille}, \citenamefont
  {R{\"o}hlsberger}, \citenamefont {Wolny},\ and\ \citenamefont
  {Sch{\"u}nemann}}]{Sakshath2017}%
  \BibitemOpen
  \bibfield  {author} {\bibinfo {author} {\bibfnamefont {S.}~\bibnamefont
  {Sakshath}}, \bibinfo {author} {\bibfnamefont {K.}~\bibnamefont {Jenni}},
  \bibinfo {author} {\bibfnamefont {L.}~\bibnamefont {Scherthan}}, \bibinfo
  {author} {\bibfnamefont {P.}~\bibnamefont {W{\"u}rtz}}, \bibinfo {author}
  {\bibfnamefont {M.}~\bibnamefont {Herlitschke}}, \bibinfo {author}
  {\bibfnamefont {I.}~\bibnamefont {Sergeev}}, \bibinfo {author} {\bibfnamefont
  {C.}~\bibnamefont {Strohm}}, \bibinfo {author} {\bibfnamefont {H.-C.}\
  \bibnamefont {Wille}}, \bibinfo {author} {\bibfnamefont {R.}~\bibnamefont
  {R{\"o}hlsberger}}, \bibinfo {author} {\bibfnamefont {J.~A.}\ \bibnamefont
  {Wolny}}, \ and\ \bibinfo {author} {\bibfnamefont {V.}~\bibnamefont
  {Sch{\"u}nemann}},\ }\bibfield  {title} {\emph {\bibinfo {title} {Optical
  pump - nuclear resonance probe experiments on spin crossover complexes},\
  }}\href {\doibase 10.1007/s10751-017-1461-3} {\bibfield  {journal} {\bibinfo
  {journal} {Hyperfine Interactions}\ }\textbf {\bibinfo {volume} {238}},\
  \bibinfo {pages} {89} (\bibinfo {year} {2017})}\BibitemShut {NoStop}%
\bibitem [{\citenamefont {Heeg}\ \emph {et~al.}(2017)\citenamefont {Heeg},
  \citenamefont {Kaldun}, \citenamefont {Strohm}, \citenamefont {Reiser},
  \citenamefont {Ott}, \citenamefont {Subramanian}, \citenamefont {Lentrodt},
  \citenamefont {Haber}, \citenamefont {Wille}, \citenamefont {Goerttler},
  \citenamefont {R{\"u}ffer}, \citenamefont {Keitel}, \citenamefont
  {R{\"o}hlsberger}, \citenamefont {Pfeifer},\ and\ \citenamefont
  {Evers}}]{Heeg2017}%
  \BibitemOpen
  \bibfield  {author} {\bibinfo {author} {\bibfnamefont {K.~P.}\ \bibnamefont
  {Heeg}}, \bibinfo {author} {\bibfnamefont {A.}~\bibnamefont {Kaldun}},
  \bibinfo {author} {\bibfnamefont {C.}~\bibnamefont {Strohm}}, \bibinfo
  {author} {\bibfnamefont {P.}~\bibnamefont {Reiser}}, \bibinfo {author}
  {\bibfnamefont {C.}~\bibnamefont {Ott}}, \bibinfo {author} {\bibfnamefont
  {R.}~\bibnamefont {Subramanian}}, \bibinfo {author} {\bibfnamefont
  {D.}~\bibnamefont {Lentrodt}}, \bibinfo {author} {\bibfnamefont
  {J.}~\bibnamefont {Haber}}, \bibinfo {author} {\bibfnamefont {H.-C.}\
  \bibnamefont {Wille}}, \bibinfo {author} {\bibfnamefont {S.}~\bibnamefont
  {Goerttler}}, \bibinfo {author} {\bibfnamefont {R.}~\bibnamefont
  {R{\"u}ffer}}, \bibinfo {author} {\bibfnamefont {C.~H.}\ \bibnamefont
  {Keitel}}, \bibinfo {author} {\bibfnamefont {R.}~\bibnamefont
  {R{\"o}hlsberger}}, \bibinfo {author} {\bibfnamefont {T.}~\bibnamefont
  {Pfeifer}}, \ and\ \bibinfo {author} {\bibfnamefont {J.}~\bibnamefont
  {Evers}},\ }\bibfield  {title} {\emph {\bibinfo {title} {Spectral narrowing
  of x-ray pulses for precision spectroscopy with nuclear resonances},\ }}\href
  {\doibase 10.1126/science.aan3512} {\bibfield  {journal} {\bibinfo  {journal}
  {Science}\ }\textbf {\bibinfo {volume} {357}},\ \bibinfo {pages} {375}
  (\bibinfo {year} {2017})}\BibitemShut {NoStop}%
\bibitem [{\citenamefont {Heeg}\ \emph {et~al.}(2021)\citenamefont {Heeg},
  \citenamefont {Kaldun}, \citenamefont {Strohm}, \citenamefont {Ott},
  \citenamefont {Subramanian}, \citenamefont {Lentrodt}, \citenamefont {Haber},
  \citenamefont {Wille}, \citenamefont {Goerttler}, \citenamefont {R{\"u}ffer},
  \citenamefont {Keitel}, \citenamefont {R{\"o}hlsberger}, \citenamefont
  {Pfeifer},\ and\ \citenamefont {Evers}}]{Heeg2021}%
  \BibitemOpen
  \bibfield  {author} {\bibinfo {author} {\bibfnamefont {K.~P.}\ \bibnamefont
  {Heeg}}, \bibinfo {author} {\bibfnamefont {A.}~\bibnamefont {Kaldun}},
  \bibinfo {author} {\bibfnamefont {C.}~\bibnamefont {Strohm}}, \bibinfo
  {author} {\bibfnamefont {C.}~\bibnamefont {Ott}}, \bibinfo {author}
  {\bibfnamefont {R.}~\bibnamefont {Subramanian}}, \bibinfo {author}
  {\bibfnamefont {D.}~\bibnamefont {Lentrodt}}, \bibinfo {author}
  {\bibfnamefont {J.}~\bibnamefont {Haber}}, \bibinfo {author} {\bibfnamefont
  {H.-C.}\ \bibnamefont {Wille}}, \bibinfo {author} {\bibfnamefont
  {S.}~\bibnamefont {Goerttler}}, \bibinfo {author} {\bibfnamefont
  {R.}~\bibnamefont {R{\"u}ffer}}, \bibinfo {author} {\bibfnamefont {C.~H.}\
  \bibnamefont {Keitel}}, \bibinfo {author} {\bibfnamefont {R.}~\bibnamefont
  {R{\"o}hlsberger}}, \bibinfo {author} {\bibfnamefont {T.}~\bibnamefont
  {Pfeifer}}, \ and\ \bibinfo {author} {\bibfnamefont {J.}~\bibnamefont
  {Evers}},\ }\bibfield  {title} {\emph {\bibinfo {title} {Coherent
  x-ray-optical control of nuclear excitons},\ }}\href {\doibase
  10.1038/s41586-021-03276-x} {\bibfield  {journal} {\bibinfo  {journal}
  {Nature}\ }\textbf {\bibinfo {volume} {590}},\ \bibinfo {pages} {401}
  (\bibinfo {year} {2021})}\BibitemShut {NoStop}%
\bibitem [{\citenamefont {Bocklage}\ \emph {et~al.}(2021)\citenamefont
  {Bocklage}, \citenamefont {Gollwitzer}, \citenamefont {Strohm}, \citenamefont
  {Adolff}, \citenamefont {Schlage}, \citenamefont {Sergeev}, \citenamefont
  {Leupold}, \citenamefont {Wille}, \citenamefont {Meier},\ and\ \citenamefont
  {R{\"o}hlsberger}}]{Bocklage2021}%
  \BibitemOpen
  \bibfield  {author} {\bibinfo {author} {\bibfnamefont {L.}~\bibnamefont
  {Bocklage}}, \bibinfo {author} {\bibfnamefont {J.}~\bibnamefont
  {Gollwitzer}}, \bibinfo {author} {\bibfnamefont {C.}~\bibnamefont {Strohm}},
  \bibinfo {author} {\bibfnamefont {C.~F.}\ \bibnamefont {Adolff}}, \bibinfo
  {author} {\bibfnamefont {K.}~\bibnamefont {Schlage}}, \bibinfo {author}
  {\bibfnamefont {I.}~\bibnamefont {Sergeev}}, \bibinfo {author} {\bibfnamefont
  {O.}~\bibnamefont {Leupold}}, \bibinfo {author} {\bibfnamefont {H.-C.}\
  \bibnamefont {Wille}}, \bibinfo {author} {\bibfnamefont {G.}~\bibnamefont
  {Meier}}, \ and\ \bibinfo {author} {\bibfnamefont {R.}~\bibnamefont
  {R{\"o}hlsberger}},\ }\bibfield  {title} {\emph {\bibinfo {title} {Coherent
  control of collective nuclear quantum states via transient magnons},\ }}\href
  {\doibase 10.1126/sciadv.abc3991} {\bibfield  {journal} {\bibinfo  {journal}
  {Science Advances}\ }\textbf {\bibinfo {volume} {7}} (\bibinfo {year}
  {2021}),\ 10.1126/sciadv.abc3991},\ \Eprint
  {http://arxiv.org/abs/https://advances.sciencemag.org/content/7/5/eabc3991.full.pdf}
  {https://advances.sciencemag.org/content/7/5/eabc3991.full.pdf} \BibitemShut
  {NoStop}%
\bibitem [{\citenamefont {R{\"{o}}hlsberger}(2005)}]{Rohlsberger2005}%
  \BibitemOpen
  \bibfield  {author} {\bibinfo {author} {\bibfnamefont {R.}~\bibnamefont
  {R{\"{o}}hlsberger}},\ }\href {\doibase 10.1007/b86125} {\emph {\bibinfo
  {title} {Nuclear Condensed Matter Physics with Synchrotron Radiation}}},\
  \bibinfo {series} {Springer Tracts in Modern Physics}, Vol.\ \bibinfo
  {volume} {208}\ (\bibinfo  {publisher} {Springer},\ \bibinfo {address}
  {Berlin, Heidelberg},\ \bibinfo {year} {2005})\BibitemShut {NoStop}%
\bibitem [{\citenamefont {Sturhahn}(2000)}]{Sturhahn2000}%
  \BibitemOpen
  \bibfield  {author} {\bibinfo {author} {\bibfnamefont {W.}~\bibnamefont
  {Sturhahn}},\ }\bibfield  {title} {\emph {\bibinfo {title} {Conuss and
  phoenix: Evaluation of nuclear resonant scattering data},\ }}\href {\doibase
  10.1023/A:1012681503686} {\bibfield  {journal} {\bibinfo  {journal}
  {Hyperfine Interactions}\ }\textbf {\bibinfo {volume} {125}},\ \bibinfo
  {pages} {149} (\bibinfo {year} {2000})}\BibitemShut {NoStop}%
\bibitem [{\citenamefont {Manzoni}\ \emph {et~al.}(2017)\citenamefont
  {Manzoni}, \citenamefont {Chang},\ and\ \citenamefont
  {Douglas}}]{Manzoni2017}%
  \BibitemOpen
  \bibfield  {author} {\bibinfo {author} {\bibfnamefont {M.~T.}\ \bibnamefont
  {Manzoni}}, \bibinfo {author} {\bibfnamefont {D.~E.}\ \bibnamefont {Chang}},
  \ and\ \bibinfo {author} {\bibfnamefont {J.~S.}\ \bibnamefont {Douglas}},\
  }\bibfield  {title} {\emph {\bibinfo {title} {Simulating quantum light
  propagation through atomic ensembles using matrix product states},\ }}\href
  {\doibase 10.1038/s41467-017-01416-4} {\bibfield  {journal} {\bibinfo
  {journal} {Nature Communications}\ }\textbf {\bibinfo {volume} {8}},\
  \bibinfo {pages} {1743} (\bibinfo {year} {2017})}\BibitemShut {NoStop}%
\bibitem [{\citenamefont {Mahmoodian}\ \emph {et~al.}(2018)\citenamefont
  {Mahmoodian}, \citenamefont {\ifmmode~\check{C}\else \v{C}\fi{}epulkovskis},
  \citenamefont {Das}, \citenamefont {Lodahl}, \citenamefont {Hammerer},\ and\
  \citenamefont {S\o{}rensen}}]{Mahmoodian2018}%
  \BibitemOpen
  \bibfield  {author} {\bibinfo {author} {\bibfnamefont {S.}~\bibnamefont
  {Mahmoodian}}, \bibinfo {author} {\bibfnamefont {M.}~\bibnamefont
  {\ifmmode~\check{C}\else \v{C}\fi{}epulkovskis}}, \bibinfo {author}
  {\bibfnamefont {S.}~\bibnamefont {Das}}, \bibinfo {author} {\bibfnamefont
  {P.}~\bibnamefont {Lodahl}}, \bibinfo {author} {\bibfnamefont
  {K.}~\bibnamefont {Hammerer}}, \ and\ \bibinfo {author} {\bibfnamefont
  {A.~S.}\ \bibnamefont {S\o{}rensen}},\ }\bibfield  {title} {\emph {\bibinfo
  {title} {Strongly correlated photon transport in waveguide quantum
  electrodynamics with weakly coupled emitters},\ }}\href {\doibase
  10.1103/PhysRevLett.121.143601} {\bibfield  {journal} {\bibinfo  {journal}
  {Phys. Rev. Lett.}\ }\textbf {\bibinfo {volume} {121}},\ \bibinfo {pages}
  {143601} (\bibinfo {year} {2018})}\BibitemShut {NoStop}%
\bibitem [{\citenamefont {Prasad}\ \emph {et~al.}(2020)\citenamefont {Prasad},
  \citenamefont {Hinney}, \citenamefont {Mahmoodian}, \citenamefont {Hammerer},
  \citenamefont {Rind}, \citenamefont {Schneeweiss}, \citenamefont
  {S{\o}rensen}, \citenamefont {Volz},\ and\ \citenamefont
  {Rauschenbeutel}}]{Prasad2020}%
  \BibitemOpen
  \bibfield  {author} {\bibinfo {author} {\bibfnamefont {A.~S.}\ \bibnamefont
  {Prasad}}, \bibinfo {author} {\bibfnamefont {J.}~\bibnamefont {Hinney}},
  \bibinfo {author} {\bibfnamefont {S.}~\bibnamefont {Mahmoodian}}, \bibinfo
  {author} {\bibfnamefont {K.}~\bibnamefont {Hammerer}}, \bibinfo {author}
  {\bibfnamefont {S.}~\bibnamefont {Rind}}, \bibinfo {author} {\bibfnamefont
  {P.}~\bibnamefont {Schneeweiss}}, \bibinfo {author} {\bibfnamefont {A.~S.}\
  \bibnamefont {S{\o}rensen}}, \bibinfo {author} {\bibfnamefont
  {J.}~\bibnamefont {Volz}}, \ and\ \bibinfo {author} {\bibfnamefont
  {A.}~\bibnamefont {Rauschenbeutel}},\ }\bibfield  {title} {\emph {\bibinfo
  {title} {Correlating photons using the collective nonlinear response of atoms
  weakly coupled to an optical mode},\ }}\href {\doibase
  10.1038/s41566-020-0692-z} {\bibfield  {journal} {\bibinfo  {journal} {Nature
  Photonics}\ }\textbf {\bibinfo {volume} {14}},\ \bibinfo {pages} {719}
  (\bibinfo {year} {2020})}\BibitemShut {NoStop}%
\bibitem [{\citenamefont {Cordier}\ \emph {et~al.}(2023)\citenamefont
  {Cordier}, \citenamefont {Schemmer}, \citenamefont {Schneeweiss},
  \citenamefont {Volz},\ and\ \citenamefont {Rauschenbeutel}}]{Cordier2023}%
  \BibitemOpen
  \bibfield  {author} {\bibinfo {author} {\bibfnamefont {M.}~\bibnamefont
  {Cordier}}, \bibinfo {author} {\bibfnamefont {M.}~\bibnamefont {Schemmer}},
  \bibinfo {author} {\bibfnamefont {P.}~\bibnamefont {Schneeweiss}}, \bibinfo
  {author} {\bibfnamefont {J.}~\bibnamefont {Volz}}, \ and\ \bibinfo {author}
  {\bibfnamefont {A.}~\bibnamefont {Rauschenbeutel}},\ }\bibfield  {title}
  {\emph {\bibinfo {title} {Tailoring photon statistics with an atom-based
  two-photon interferometer},\ }}\href {\doibase
  10.1103/PhysRevLett.131.183601} {\bibfield  {journal} {\bibinfo  {journal}
  {Phys. Rev. Lett.}\ }\textbf {\bibinfo {volume} {131}},\ \bibinfo {pages}
  {183601} (\bibinfo {year} {2023})}\BibitemShut {NoStop}%
\bibitem [{\citenamefont {Liedl}\ \emph {et~al.}(2024)\citenamefont {Liedl},
  \citenamefont {Tebbenjohanns}, \citenamefont {Bach}, \citenamefont {Pucher},
  \citenamefont {Rauschenbeutel},\ and\ \citenamefont
  {Schneeweiss}}]{Liedl2024}%
  \BibitemOpen
  \bibfield  {author} {\bibinfo {author} {\bibfnamefont {C.}~\bibnamefont
  {Liedl}}, \bibinfo {author} {\bibfnamefont {F.}~\bibnamefont
  {Tebbenjohanns}}, \bibinfo {author} {\bibfnamefont {C.}~\bibnamefont {Bach}},
  \bibinfo {author} {\bibfnamefont {S.}~\bibnamefont {Pucher}}, \bibinfo
  {author} {\bibfnamefont {A.}~\bibnamefont {Rauschenbeutel}}, \ and\ \bibinfo
  {author} {\bibfnamefont {P.}~\bibnamefont {Schneeweiss}},\ }\bibfield
  {title} {\emph {\bibinfo {title} {Observation of superradiant bursts in a
  cascaded quantum system},\ }}\href {\doibase 10.1103/PhysRevX.14.011020}
  {\bibfield  {journal} {\bibinfo  {journal} {Phys. Rev. X}\ }\textbf {\bibinfo
  {volume} {14}},\ \bibinfo {pages} {011020} (\bibinfo {year}
  {2024})}\BibitemShut {NoStop}%
\bibitem [{\citenamefont {Chumakov}\ \emph {et~al.}(2018)\citenamefont
  {Chumakov}, \citenamefont {Baron}, \citenamefont {Sergueev}, \citenamefont
  {Strohm}, \citenamefont {Leupold}, \citenamefont {Shvyd'ko}, \citenamefont
  {Smirnov}, \citenamefont {R{\"u}ffer}, \citenamefont {Inubushi},
  \citenamefont {Yabashi}, \citenamefont {Tono}, \citenamefont {Kudo},\ and\
  \citenamefont {Ishikawa}}]{Chumakov2018}%
  \BibitemOpen
  \bibfield  {author} {\bibinfo {author} {\bibfnamefont {A.~I.}\ \bibnamefont
  {Chumakov}}, \bibinfo {author} {\bibfnamefont {A.~Q.~R.}\ \bibnamefont
  {Baron}}, \bibinfo {author} {\bibfnamefont {I.}~\bibnamefont {Sergueev}},
  \bibinfo {author} {\bibfnamefont {C.}~\bibnamefont {Strohm}}, \bibinfo
  {author} {\bibfnamefont {O.}~\bibnamefont {Leupold}}, \bibinfo {author}
  {\bibfnamefont {Y.}~\bibnamefont {Shvyd'ko}}, \bibinfo {author}
  {\bibfnamefont {G.~V.}\ \bibnamefont {Smirnov}}, \bibinfo {author}
  {\bibfnamefont {R.}~\bibnamefont {R{\"u}ffer}}, \bibinfo {author}
  {\bibfnamefont {Y.}~\bibnamefont {Inubushi}}, \bibinfo {author}
  {\bibfnamefont {M.}~\bibnamefont {Yabashi}}, \bibinfo {author} {\bibfnamefont
  {K.}~\bibnamefont {Tono}}, \bibinfo {author} {\bibfnamefont {T.}~\bibnamefont
  {Kudo}}, \ and\ \bibinfo {author} {\bibfnamefont {T.}~\bibnamefont
  {Ishikawa}},\ }\bibfield  {title} {\emph {\bibinfo {title} {Superradiance of
  an ensemble of nuclei excited by a free electron laser},\ }}\href {\doibase
  10.1038/s41567-017-0001-z} {\bibfield  {journal} {\bibinfo  {journal} {Nat.
  Phys.}\ }\textbf {\bibinfo {volume} {14}},\ \bibinfo {pages} {261} (\bibinfo
  {year} {2018})}\BibitemShut {NoStop}%
\bibitem [{\citenamefont {Shvyd'ko}\ \emph {et~al.}(2023)\citenamefont
  {Shvyd'ko}, \citenamefont {R{\"o}hlsberger}, \citenamefont {Kocharovskaya},
  \citenamefont {Evers}, \citenamefont {Geloni}, \citenamefont {Liu},
  \citenamefont {Shu}, \citenamefont {Miceli}, \citenamefont {Stone},
  \citenamefont {Hippler}, \citenamefont {Marx-Glowna}, \citenamefont
  {Uschmann}, \citenamefont {Loetzsch}, \citenamefont {Leupold}, \citenamefont
  {Wille}, \citenamefont {Sergeev}, \citenamefont {Gerharz}, \citenamefont
  {Zhang}, \citenamefont {Grech}, \citenamefont {Guetg}, \citenamefont
  {Kocharyan}, \citenamefont {Kujala}, \citenamefont {Liu}, \citenamefont
  {Qin}, \citenamefont {Zozulya}, \citenamefont {Hallmann}, \citenamefont
  {Boesenberg}, \citenamefont {Jo}, \citenamefont {M{\"o}ller}, \citenamefont
  {Rodriguez-Fernandez}, \citenamefont {Youssef}, \citenamefont {Madsen},\ and\
  \citenamefont {Kolodziej}}]{Shvydko2023}%
  \BibitemOpen
  \bibfield  {author} {\bibinfo {author} {\bibfnamefont {Y.}~\bibnamefont
  {Shvyd'ko}}, \bibinfo {author} {\bibfnamefont {R.}~\bibnamefont
  {R{\"o}hlsberger}}, \bibinfo {author} {\bibfnamefont {O.}~\bibnamefont
  {Kocharovskaya}}, \bibinfo {author} {\bibfnamefont {J.}~\bibnamefont
  {Evers}}, \bibinfo {author} {\bibfnamefont {G.~A.}\ \bibnamefont {Geloni}},
  \bibinfo {author} {\bibfnamefont {P.}~\bibnamefont {Liu}}, \bibinfo {author}
  {\bibfnamefont {D.}~\bibnamefont {Shu}}, \bibinfo {author} {\bibfnamefont
  {A.}~\bibnamefont {Miceli}}, \bibinfo {author} {\bibfnamefont
  {B.}~\bibnamefont {Stone}}, \bibinfo {author} {\bibfnamefont
  {W.}~\bibnamefont {Hippler}}, \bibinfo {author} {\bibfnamefont
  {B.}~\bibnamefont {Marx-Glowna}}, \bibinfo {author} {\bibfnamefont
  {I.}~\bibnamefont {Uschmann}}, \bibinfo {author} {\bibfnamefont
  {R.}~\bibnamefont {Loetzsch}}, \bibinfo {author} {\bibfnamefont
  {O.}~\bibnamefont {Leupold}}, \bibinfo {author} {\bibfnamefont {H.-C.}\
  \bibnamefont {Wille}}, \bibinfo {author} {\bibfnamefont {I.}~\bibnamefont
  {Sergeev}}, \bibinfo {author} {\bibfnamefont {M.}~\bibnamefont {Gerharz}},
  \bibinfo {author} {\bibfnamefont {X.}~\bibnamefont {Zhang}}, \bibinfo
  {author} {\bibfnamefont {C.}~\bibnamefont {Grech}}, \bibinfo {author}
  {\bibfnamefont {M.}~\bibnamefont {Guetg}}, \bibinfo {author} {\bibfnamefont
  {V.}~\bibnamefont {Kocharyan}}, \bibinfo {author} {\bibfnamefont
  {N.}~\bibnamefont {Kujala}}, \bibinfo {author} {\bibfnamefont
  {S.}~\bibnamefont {Liu}}, \bibinfo {author} {\bibfnamefont {W.}~\bibnamefont
  {Qin}}, \bibinfo {author} {\bibfnamefont {A.}~\bibnamefont {Zozulya}},
  \bibinfo {author} {\bibfnamefont {J.}~\bibnamefont {Hallmann}}, \bibinfo
  {author} {\bibfnamefont {U.}~\bibnamefont {Boesenberg}}, \bibinfo {author}
  {\bibfnamefont {W.}~\bibnamefont {Jo}}, \bibinfo {author} {\bibfnamefont
  {J.}~\bibnamefont {M{\"o}ller}}, \bibinfo {author} {\bibfnamefont
  {A.}~\bibnamefont {Rodriguez-Fernandez}}, \bibinfo {author} {\bibfnamefont
  {M.}~\bibnamefont {Youssef}}, \bibinfo {author} {\bibfnamefont
  {A.}~\bibnamefont {Madsen}}, \ and\ \bibinfo {author} {\bibfnamefont
  {T.}~\bibnamefont {Kolodziej}},\ }\bibfield  {title} {\emph {\bibinfo {title}
  {Resonant x-ray excitation of the nuclear clock isomer 45sc},\ }}\href
  {\doibase 10.1038/s41586-023-06491-w} {\bibfield  {journal} {\bibinfo
  {journal} {Nature}\ }\textbf {\bibinfo {volume} {622}},\ \bibinfo {pages}
  {471} (\bibinfo {year} {2023})}\BibitemShut {NoStop}%
\bibitem [{\citenamefont {Matinyan}(1998)}]{Matinyan1998}%
  \BibitemOpen
  \bibfield  {author} {\bibinfo {author} {\bibfnamefont {S.}~\bibnamefont
  {Matinyan}},\ }\bibfield  {title} {\emph {\bibinfo {title} {Lasers as a
  bridge between atomic and nuclear physics},\ }}\href {\doibase
  https://doi.org/10.1016/S0370-1573(97)00084-7} {\bibfield  {journal}
  {\bibinfo  {journal} {Physics Reports}\ }\textbf {\bibinfo {volume} {298}},\
  \bibinfo {pages} {199} (\bibinfo {year} {1998})}\BibitemShut {NoStop}%
\bibitem [{\citenamefont {B\"urvenich}\ \emph {et~al.}(2006)\citenamefont
  {B\"urvenich}, \citenamefont {Evers},\ and\ \citenamefont
  {Keitel}}]{Buervenich2006}%
  \BibitemOpen
  \bibfield  {author} {\bibinfo {author} {\bibfnamefont {T.~J.}\ \bibnamefont
  {B\"urvenich}}, \bibinfo {author} {\bibfnamefont {J.}~\bibnamefont {Evers}},
  \ and\ \bibinfo {author} {\bibfnamefont {C.~H.}\ \bibnamefont {Keitel}},\
  }\bibfield  {title} {\emph {\bibinfo {title} {Nuclear quantum optics with
  x-ray laser pulses},\ }}\href {\doibase 10.1103/PhysRevLett.96.142501}
  {\bibfield  {journal} {\bibinfo  {journal} {Phys. Rev. Lett.}\ }\textbf
  {\bibinfo {volume} {96}},\ \bibinfo {pages} {142501} (\bibinfo {year}
  {2006})}\BibitemShut {NoStop}%
\bibitem [{\citenamefont {P{\'a}lffy}\ \emph {et~al.}(2008)\citenamefont
  {P{\'a}lffy}, \citenamefont {Evers},\ and\ \citenamefont
  {Keitel}}]{Palffy2008}%
  \BibitemOpen
  \bibfield  {author} {\bibinfo {author} {\bibfnamefont {A.}~\bibnamefont
  {P{\'a}lffy}}, \bibinfo {author} {\bibfnamefont {J.}~\bibnamefont {Evers}}, \
  and\ \bibinfo {author} {\bibfnamefont {C.~H.}\ \bibnamefont {Keitel}},\
  }\bibfield  {title} {\emph {\bibinfo {title} {Electric-dipole-forbidden
  nuclear transitions driven by super-intense laser fields},\ }}\href {\doibase
  10.1103/PhysRevC.77.044602} {\bibfield  {journal} {\bibinfo  {journal} {Phys.
  Rev. C}\ }\textbf {\bibinfo {volume} {77}},\ \bibinfo {pages} {044602}
  (\bibinfo {year} {2008})}\BibitemShut {NoStop}%
\bibitem [{\citenamefont {Heeg}(2014)}]{HeegPhD}%
  \BibitemOpen
  \bibfield  {author} {\bibinfo {author} {\bibfnamefont {K.~P.}\ \bibnamefont
  {Heeg}},\ }\href {\doibase 10.11588/heidok.00017869} {\bibinfo {title} {X-ray
  quantum optics with {M}{\"o}ssbauer nuclei in thin-film cavities},\ }
  (\bibinfo {year} {2014})\BibitemShut {NoStop}%
\bibitem [{\citenamefont {Junker}\ \emph {et~al.}(2012)\citenamefont {Junker},
  \citenamefont {P{\'{a}}lffy},\ and\ \citenamefont {Keitel}}]{Junker2012}%
  \BibitemOpen
  \bibfield  {author} {\bibinfo {author} {\bibfnamefont {A.}~\bibnamefont
  {Junker}}, \bibinfo {author} {\bibfnamefont {A.}~\bibnamefont
  {P{\'{a}}lffy}}, \ and\ \bibinfo {author} {\bibfnamefont {C.~H.}\
  \bibnamefont {Keitel}},\ }\bibfield  {title} {\emph {\bibinfo {title}
  {Cooperative effects in nuclear excitation with coherent x-ray light},\
  }}\href {\doibase 10.1088/1367-2630/14/8/085025} {\bibfield  {journal}
  {\bibinfo  {journal} {New Journal of Physics}\ }\textbf {\bibinfo {volume}
  {14}},\ \bibinfo {pages} {085025} (\bibinfo {year} {2012})}\BibitemShut
  {NoStop}%
\bibitem [{\citenamefont {Salditt}\ and\ \citenamefont
  {Osterhoff}(2020)}]{Salditt2020}%
  \BibitemOpen
  \bibfield  {author} {\bibinfo {author} {\bibfnamefont {T.}~\bibnamefont
  {Salditt}}\ and\ \bibinfo {author} {\bibfnamefont {M.}~\bibnamefont
  {Osterhoff}},\ }\bibinfo {title} {X-ray focusing and optics},\ in\ \href
  {\doibase 10.1007/978-3-030-34413-9_3} {\emph {\bibinfo {booktitle}
  {Nanoscale Photonic Imaging}}},\ \bibinfo {editor} {edited by\ \bibinfo
  {editor} {\bibfnamefont {T.}~\bibnamefont {Salditt}}, \bibinfo {editor}
  {\bibfnamefont {A.}~\bibnamefont {Egner}}, \ and\ \bibinfo {editor}
  {\bibfnamefont {D.~R.}\ \bibnamefont {Luke}}}\ (\bibinfo  {publisher}
  {Springer International Publishing},\ \bibinfo {address} {Cham},\ \bibinfo
  {year} {2020})\ pp.\ \bibinfo {pages} {71--124}\BibitemShut {NoStop}%
\bibitem [{\citenamefont {Schenk}(2011)}]{Schenk2011PhD}%
  \BibitemOpen
  \bibfield  {author} {\bibinfo {author} {\bibfnamefont {F.}~\bibnamefont
  {Schenk}},\ }\href {\doibase 10.17875/gup2011-75} {\emph {\bibinfo {title}
  {Optimization of resonances for multilayer x-ray resonators}}},\ \bibinfo
  {series} {Göttingen Series in X-ray Physics}, Vol.\ \bibinfo {volume} {003}\
  (\bibinfo  {publisher} {Universit{\"a}tsverlag G{\"o}ttingen},\ \bibinfo
  {address} {G{\"o}ttingen},\ \bibinfo {year} {2011})\BibitemShut {NoStop}%
\bibitem [{\citenamefont {Lohse}\ \emph {et~al.}(2024)\citenamefont {Lohse},
  \citenamefont {Andrejić}, \citenamefont {Velten}, \citenamefont {Vassholz},
  \citenamefont {Neuhaus}, \citenamefont {Negi}, \citenamefont {Panchwanee},
  \citenamefont {Sergeev}, \citenamefont {Pálffy}, \citenamefont {Salditt},\
  and\ \citenamefont {Röhlsberger}}]{Lohse2024_arxiv}%
  \BibitemOpen
  \bibfield  {author} {\bibinfo {author} {\bibfnamefont {L.~M.}\ \bibnamefont
  {Lohse}}, \bibinfo {author} {\bibfnamefont {P.}~\bibnamefont {Andrejić}},
  \bibinfo {author} {\bibfnamefont {S.}~\bibnamefont {Velten}}, \bibinfo
  {author} {\bibfnamefont {M.}~\bibnamefont {Vassholz}}, \bibinfo {author}
  {\bibfnamefont {C.}~\bibnamefont {Neuhaus}}, \bibinfo {author} {\bibfnamefont
  {A.}~\bibnamefont {Negi}}, \bibinfo {author} {\bibfnamefont {A.}~\bibnamefont
  {Panchwanee}}, \bibinfo {author} {\bibfnamefont {I.}~\bibnamefont {Sergeev}},
  \bibinfo {author} {\bibfnamefont {A.}~\bibnamefont {Pálffy}}, \bibinfo
  {author} {\bibfnamefont {T.}~\bibnamefont {Salditt}}, \ and\ \bibinfo
  {author} {\bibfnamefont {R.}~\bibnamefont {Röhlsberger}},\ }\href@noop {}
  {\bibinfo {title} {Collective nuclear excitation dynamics in mono-modal x-ray
  waveguides},\ } (\bibinfo {year} {2024}),\ \Eprint
  {http://arxiv.org/abs/2403.06508} {arXiv:2403.06508 [quant-ph]} \BibitemShut
  {NoStop}%
\bibitem [{\citenamefont {Lohse}\ and\ \citenamefont
  {Andreji{\'{c}}}(2024)}]{Lohse2024a}%
  \BibitemOpen
  \bibfield  {author} {\bibinfo {author} {\bibfnamefont {L.~M.}\ \bibnamefont
  {Lohse}}\ and\ \bibinfo {author} {\bibfnamefont {P.}~\bibnamefont
  {Andreji{\'{c}}}},\ }\bibfield  {title} {\emph {\bibinfo {title}
  {Nano-optical theory of planar x-ray waveguides},\ }}\href {\doibase
  10.1364/OE.504206} {\bibfield  {journal} {\bibinfo  {journal} {Optics
  Express}\ }\textbf {\bibinfo {volume} {32}},\ \bibinfo {pages} {9518}
  (\bibinfo {year} {2024})}\BibitemShut {NoStop}%
\bibitem [{\citenamefont {Chen}\ \emph {et~al.}(2015)\citenamefont {Chen},
  \citenamefont {Hoffmann},\ and\ \citenamefont {Salditt}}]{Chen2015}%
  \BibitemOpen
  \bibfield  {author} {\bibinfo {author} {\bibfnamefont {H.-Y.}\ \bibnamefont
  {Chen}}, \bibinfo {author} {\bibfnamefont {S.}~\bibnamefont {Hoffmann}}, \
  and\ \bibinfo {author} {\bibfnamefont {T.}~\bibnamefont {Salditt}},\
  }\bibfield  {title} {\emph {\bibinfo {title} {X-ray beam compression by
  tapered waveguides},\ }}\href {\doibase 10.1063/1.4921095} {\bibfield
  {journal} {\bibinfo  {journal} {Applied Physics Letters}\ }\textbf {\bibinfo
  {volume} {106}},\ \bibinfo {pages} {194105} (\bibinfo {year}
  {2015})}\BibitemShut {NoStop}%
\bibitem [{\citenamefont {Chen}\ \emph {et~al.}(2022)\citenamefont {Chen},
  \citenamefont {Lin}, \citenamefont {Wang}, \citenamefont {P\'alffy},\ and\
  \citenamefont {Liao}}]{Chen2022}%
  \BibitemOpen
  \bibfield  {author} {\bibinfo {author} {\bibfnamefont {Y.-H.}\ \bibnamefont
  {Chen}}, \bibinfo {author} {\bibfnamefont {P.-H.}\ \bibnamefont {Lin}},
  \bibinfo {author} {\bibfnamefont {G.-Y.}\ \bibnamefont {Wang}}, \bibinfo
  {author} {\bibfnamefont {A.}~\bibnamefont {P\'alffy}}, \ and\ \bibinfo
  {author} {\bibfnamefont {W.-T.}\ \bibnamefont {Liao}},\ }\bibfield  {title}
  {\emph {\bibinfo {title} {Transient nuclear inversion by x-ray free electron
  laser in a tapered x-ray waveguide},\ }}\href {\doibase
  10.1103/PhysRevResearch.4.L032007} {\bibfield  {journal} {\bibinfo  {journal}
  {Phys. Rev. Res.}\ }\textbf {\bibinfo {volume} {4}},\ \bibinfo {pages}
  {L032007} (\bibinfo {year} {2022})}\BibitemShut {NoStop}%
\bibitem [{\citenamefont {R{\"o}hlsberger}\ \emph {et~al.}(2014)\citenamefont
  {R{\"o}hlsberger}, \citenamefont {Evers},\ and\ \citenamefont
  {Shwartz}}]{Rohlsberger2014}%
  \BibitemOpen
  \bibfield  {author} {\bibinfo {author} {\bibfnamefont {R.}~\bibnamefont
  {R{\"o}hlsberger}}, \bibinfo {author} {\bibfnamefont {J.}~\bibnamefont
  {Evers}}, \ and\ \bibinfo {author} {\bibfnamefont {S.}~\bibnamefont
  {Shwartz}},\ }\bibinfo {title} {Quantum and nonlinear optics with hard
  x-rays},\ in\ \href {\doibase 10.1007/978-3-319-04507-8_32-1} {\emph
  {\bibinfo {booktitle} {Synchrotron Light Sources and Free-Electron Lasers:
  Accelerator Physics, Instrumentation and Science Applications}}},\ \bibinfo
  {editor} {edited by\ \bibinfo {editor} {\bibfnamefont {E.}~\bibnamefont
  {Jaeschke}}, \bibinfo {editor} {\bibfnamefont {S.}~\bibnamefont {Khan}},
  \bibinfo {editor} {\bibfnamefont {J.~R.}\ \bibnamefont {Schneider}}, \ and\
  \bibinfo {editor} {\bibfnamefont {J.~B.}\ \bibnamefont {Hastings}}}\
  (\bibinfo  {publisher} {Springer International Publishing},\ \bibinfo
  {address} {Cham},\ \bibinfo {year} {2014})\ pp.\ \bibinfo {pages}
  {1--28}\BibitemShut {NoStop}%
\bibitem [{\citenamefont {R{\"o}hlsberger}\ and\ \citenamefont
  {Evers}(2021)}]{Rohlsberger2021}%
  \BibitemOpen
  \bibfield  {author} {\bibinfo {author} {\bibfnamefont {R.}~\bibnamefont
  {R{\"o}hlsberger}}\ and\ \bibinfo {author} {\bibfnamefont {J.}~\bibnamefont
  {Evers}},\ }\bibinfo {title} {Quantum optical phenomena in nuclear resonant
  scattering},\ in\ \href {\doibase 10.1007/978-981-15-9422-9_3} {\emph
  {\bibinfo {booktitle} {Modern {M}{\"o}ssbauer Spectroscopy}}},\ \bibinfo
  {editor} {edited by\ \bibinfo {editor} {\bibfnamefont {Y.}~\bibnamefont
  {Yoshida}}\ and\ \bibinfo {editor} {\bibfnamefont {G.}~\bibnamefont
  {Langouche}}}\ (\bibinfo  {publisher} {Springer Singapore},\ \bibinfo
  {address} {Singapore},\ \bibinfo {year} {2021})\ pp.\ \bibinfo {pages}
  {105--171}\BibitemShut {NoStop}%
\bibitem [{\citenamefont {R{\"{o}}hlsberger}\ \emph {et~al.}(2005)\citenamefont
  {R{\"{o}}hlsberger}, \citenamefont {Schlage}, \citenamefont {Klein},\ and\
  \citenamefont {Leupold}}]{Rohlsberger2005a}%
  \BibitemOpen
  \bibfield  {author} {\bibinfo {author} {\bibfnamefont {R.}~\bibnamefont
  {R{\"{o}}hlsberger}}, \bibinfo {author} {\bibfnamefont {K.}~\bibnamefont
  {Schlage}}, \bibinfo {author} {\bibfnamefont {T.}~\bibnamefont {Klein}}, \
  and\ \bibinfo {author} {\bibfnamefont {O.}~\bibnamefont {Leupold}},\
  }\bibfield  {title} {\emph {\bibinfo {title} {Accelerating the spontaneous
  emission of x rays from atoms in a cavity},\ }}\href {\doibase
  10.1103/PhysRevLett.95.097601} {\bibfield  {journal} {\bibinfo  {journal}
  {Phys. Rev. Lett.}\ }\textbf {\bibinfo {volume} {95}},\ \bibinfo {pages}
  {097601} (\bibinfo {year} {2005})}\BibitemShut {NoStop}%
\bibitem [{\citenamefont {Heeg}\ and\ \citenamefont {Evers}(2013)}]{Heeg2013b}%
  \BibitemOpen
  \bibfield  {author} {\bibinfo {author} {\bibfnamefont {K.~P.}\ \bibnamefont
  {Heeg}}\ and\ \bibinfo {author} {\bibfnamefont {J.}~\bibnamefont {Evers}},\
  }\bibfield  {title} {\emph {\bibinfo {title} {X-ray quantum optics with
  {M}\"ossbauer nuclei embedded in thin-film cavities},\ }}\href {\doibase
  10.1103/PhysRevA.88.043828} {\bibfield  {journal} {\bibinfo  {journal} {Phys.
  Rev. A}\ }\textbf {\bibinfo {volume} {88}},\ \bibinfo {pages} {043828}
  (\bibinfo {year} {2013})}\BibitemShut {NoStop}%
\bibitem [{\citenamefont {Lentrodt}\ \emph {et~al.}(2020)\citenamefont
  {Lentrodt}, \citenamefont {Heeg}, \citenamefont {Keitel},\ and\ \citenamefont
  {Evers}}]{Lentrodt2020a}%
  \BibitemOpen
  \bibfield  {author} {\bibinfo {author} {\bibfnamefont {D.}~\bibnamefont
  {Lentrodt}}, \bibinfo {author} {\bibfnamefont {K.~P.}\ \bibnamefont {Heeg}},
  \bibinfo {author} {\bibfnamefont {C.~H.}\ \bibnamefont {Keitel}}, \ and\
  \bibinfo {author} {\bibfnamefont {J.}~\bibnamefont {Evers}},\ }\bibfield
  {title} {\emph {\bibinfo {title} {Ab initio quantum models for thin-film
  x-ray cavity {QED}},\ }}\href {\doibase 10.1103/PhysRevResearch.2.023396}
  {\bibfield  {journal} {\bibinfo  {journal} {Phys. Rev. Research}\ }\textbf
  {\bibinfo {volume} {2}},\ \bibinfo {pages} {023396} (\bibinfo {year}
  {2020})}\BibitemShut {NoStop}%
\bibitem [{\citenamefont {Diekmann}\ \emph {et~al.}(2022)\citenamefont
  {Diekmann}, \citenamefont {Lentrodt},\ and\ \citenamefont
  {Evers}}]{Diekmann2022}%
  \BibitemOpen
  \bibfield  {author} {\bibinfo {author} {\bibfnamefont {O.}~\bibnamefont
  {Diekmann}}, \bibinfo {author} {\bibfnamefont {D.}~\bibnamefont {Lentrodt}},
  \ and\ \bibinfo {author} {\bibfnamefont {J.}~\bibnamefont {Evers}},\
  }\bibfield  {title} {\emph {\bibinfo {title} {Inverse design approach to
  x-ray quantum optics with m\"ossbauer nuclei in thin-film cavities},\ }}\href
  {\doibase 10.1103/PhysRevA.105.013715} {\bibfield  {journal} {\bibinfo
  {journal} {Phys. Rev. A}\ }\textbf {\bibinfo {volume} {105}},\ \bibinfo
  {pages} {013715} (\bibinfo {year} {2022})}\BibitemShut {NoStop}%
\bibitem [{\citenamefont {Heeg}\ \emph {et~al.}(2016)\citenamefont {Heeg},
  \citenamefont {Keitel},\ and\ \citenamefont {Evers}}]{Heeg2016arxiv}%
  \BibitemOpen
  \bibfield  {author} {\bibinfo {author} {\bibfnamefont {K.~P.}\ \bibnamefont
  {Heeg}}, \bibinfo {author} {\bibfnamefont {C.~H.}\ \bibnamefont {Keitel}}, \
  and\ \bibinfo {author} {\bibfnamefont {J.}~\bibnamefont {Evers}},\ }\bibfield
   {title} {\emph {\bibinfo {title} {Inducing and detecting collective
  population inversions of m\"ossbauer nuclei},\ }}\href@noop {} {\  (\bibinfo
  {year} {2016})},\ \Eprint {http://arxiv.org/abs/1607.04116} {arXiv:1607.04116
  [quant-ph]} \BibitemShut {NoStop}%
\bibitem [{com()}]{companion_nonlin2024_arxivMethod}%
  \BibitemOpen
  \href@noop {} {}\bibinfo {howpublished} {See Companion paper at
  \url{https://arxiv.org/abs/2405.12780}.}\BibitemShut {Stop}%
\bibitem [{\citenamefont {Asenjo-Garcia}\ \emph {et~al.}(2017)\citenamefont
  {Asenjo-Garcia}, \citenamefont {Hood}, \citenamefont {Chang},\ and\
  \citenamefont {Kimble}}]{Asenjo-Garcia2017a}%
  \BibitemOpen
  \bibfield  {author} {\bibinfo {author} {\bibfnamefont {A.}~\bibnamefont
  {Asenjo-Garcia}}, \bibinfo {author} {\bibfnamefont {J.~D.}\ \bibnamefont
  {Hood}}, \bibinfo {author} {\bibfnamefont {D.~E.}\ \bibnamefont {Chang}}, \
  and\ \bibinfo {author} {\bibfnamefont {H.~J.}\ \bibnamefont {Kimble}},\
  }\bibfield  {title} {\emph {\bibinfo {title} {Atom-light interactions in
  quasi-one-dimensional nanostructures: A {G}reen's-function perspective},\
  }}\href {\doibase 10.1103/PhysRevA.95.033818} {\bibfield  {journal} {\bibinfo
   {journal} {Phys. Rev. A}\ }\textbf {\bibinfo {volume} {95}},\ \bibinfo
  {pages} {033818} (\bibinfo {year} {2017})}\BibitemShut {NoStop}%
\bibitem [{\citenamefont {Mink}\ and\ \citenamefont
  {Fleischhauer}(2023)}]{Mink2023}%
  \BibitemOpen
  \bibfield  {author} {\bibinfo {author} {\bibfnamefont {C.~D.}\ \bibnamefont
  {Mink}}\ and\ \bibinfo {author} {\bibfnamefont {M.}~\bibnamefont
  {Fleischhauer}},\ }\bibfield  {title} {\emph {\bibinfo {title} {{Collective
  radiative interactions in the discrete truncated Wigner approximation}},\
  }}\href {\doibase 10.21468/SciPostPhys.15.6.233} {\bibfield  {journal}
  {\bibinfo  {journal} {SciPost Phys.}\ }\textbf {\bibinfo {volume} {15}},\
  \bibinfo {pages} {233} (\bibinfo {year} {2023})}\BibitemShut {NoStop}%
\bibitem [{\citenamefont {Schachenmayer}\ \emph {et~al.}(2015)\citenamefont
  {Schachenmayer}, \citenamefont {Pikovski},\ and\ \citenamefont
  {Rey}}]{Schachenmayer2015}%
  \BibitemOpen
  \bibfield  {author} {\bibinfo {author} {\bibfnamefont {J.}~\bibnamefont
  {Schachenmayer}}, \bibinfo {author} {\bibfnamefont {A.}~\bibnamefont
  {Pikovski}}, \ and\ \bibinfo {author} {\bibfnamefont {A.~M.}\ \bibnamefont
  {Rey}},\ }\bibfield  {title} {\emph {\bibinfo {title} {Many-body quantum spin
  dynamics with monte carlo trajectories on a discrete phase space},\ }}\href
  {\doibase 10.1103/PhysRevX.5.011022} {\bibfield  {journal} {\bibinfo
  {journal} {Phys. Rev. X}\ }\textbf {\bibinfo {volume} {5}},\ \bibinfo {pages}
  {011022} (\bibinfo {year} {2015})}\BibitemShut {NoStop}%
\bibitem [{\citenamefont {Moreno-Cardoner}\ \emph {et~al.}(2021)\citenamefont
  {Moreno-Cardoner}, \citenamefont {Goncalves},\ and\ \citenamefont
  {Chang}}]{Moreno-Cardoner2021}%
  \BibitemOpen
  \bibfield  {author} {\bibinfo {author} {\bibfnamefont {M.}~\bibnamefont
  {Moreno-Cardoner}}, \bibinfo {author} {\bibfnamefont {D.}~\bibnamefont
  {Goncalves}}, \ and\ \bibinfo {author} {\bibfnamefont {D.~E.}\ \bibnamefont
  {Chang}},\ }\bibfield  {title} {\emph {\bibinfo {title} {Quantum nonlinear
  optics based on two-dimensional rydberg atom arrays},\ }}\href {\doibase
  10.1103/PhysRevLett.127.263602} {\bibfield  {journal} {\bibinfo  {journal}
  {Phys. Rev. Lett.}\ }\textbf {\bibinfo {volume} {127}},\ \bibinfo {pages}
  {263602} (\bibinfo {year} {2021})}\BibitemShut {NoStop}%
\bibitem [{\citenamefont {Amann}\ \emph {et~al.}(2012)\citenamefont {Amann},
  \citenamefont {Berg}, \citenamefont {Blank}, \citenamefont {Decker},
  \citenamefont {Ding}, \citenamefont {Emma}, \citenamefont {Feng},
  \citenamefont {Frisch}, \citenamefont {Fritz}, \citenamefont {Hastings},
  \citenamefont {Huang}, \citenamefont {Krzywinski}, \citenamefont {Lindberg},
  \citenamefont {Loos}, \citenamefont {Lutman}, \citenamefont {Nuhn},
  \citenamefont {Ratner}, \citenamefont {Rzepiela}, \citenamefont {Shu},
  \citenamefont {Shvyd'ko}, \citenamefont {Spampinati}, \citenamefont
  {Stoupin}, \citenamefont {Terentyev}, \citenamefont {Trakhtenberg},
  \citenamefont {Walz}, \citenamefont {Welch}, \citenamefont {Wu},
  \citenamefont {Zholents},\ and\ \citenamefont {Zhu}}]{Amann2012}%
  \BibitemOpen
  \bibfield  {author} {\bibinfo {author} {\bibfnamefont {J.}~\bibnamefont
  {Amann}}, \bibinfo {author} {\bibfnamefont {W.}~\bibnamefont {Berg}},
  \bibinfo {author} {\bibfnamefont {V.}~\bibnamefont {Blank}}, \bibinfo
  {author} {\bibfnamefont {F.-J.}\ \bibnamefont {Decker}}, \bibinfo {author}
  {\bibfnamefont {Y.}~\bibnamefont {Ding}}, \bibinfo {author} {\bibfnamefont
  {P.}~\bibnamefont {Emma}}, \bibinfo {author} {\bibfnamefont {Y.}~\bibnamefont
  {Feng}}, \bibinfo {author} {\bibfnamefont {J.}~\bibnamefont {Frisch}},
  \bibinfo {author} {\bibfnamefont {D.}~\bibnamefont {Fritz}}, \bibinfo
  {author} {\bibfnamefont {J.}~\bibnamefont {Hastings}}, \bibinfo {author}
  {\bibfnamefont {Z.}~\bibnamefont {Huang}}, \bibinfo {author} {\bibfnamefont
  {J.}~\bibnamefont {Krzywinski}}, \bibinfo {author} {\bibfnamefont
  {R.}~\bibnamefont {Lindberg}}, \bibinfo {author} {\bibfnamefont
  {H.}~\bibnamefont {Loos}}, \bibinfo {author} {\bibfnamefont {A.}~\bibnamefont
  {Lutman}}, \bibinfo {author} {\bibfnamefont {H.-D.}\ \bibnamefont {Nuhn}},
  \bibinfo {author} {\bibfnamefont {D.}~\bibnamefont {Ratner}}, \bibinfo
  {author} {\bibfnamefont {J.}~\bibnamefont {Rzepiela}}, \bibinfo {author}
  {\bibfnamefont {D.}~\bibnamefont {Shu}}, \bibinfo {author} {\bibfnamefont
  {Y.}~\bibnamefont {Shvyd'ko}}, \bibinfo {author} {\bibfnamefont
  {S.}~\bibnamefont {Spampinati}}, \bibinfo {author} {\bibfnamefont
  {S.}~\bibnamefont {Stoupin}}, \bibinfo {author} {\bibfnamefont
  {S.}~\bibnamefont {Terentyev}}, \bibinfo {author} {\bibfnamefont
  {E.}~\bibnamefont {Trakhtenberg}}, \bibinfo {author} {\bibfnamefont
  {D.}~\bibnamefont {Walz}}, \bibinfo {author} {\bibfnamefont {J.}~\bibnamefont
  {Welch}}, \bibinfo {author} {\bibfnamefont {J.}~\bibnamefont {Wu}}, \bibinfo
  {author} {\bibfnamefont {A.}~\bibnamefont {Zholents}}, \ and\ \bibinfo
  {author} {\bibfnamefont {D.}~\bibnamefont {Zhu}},\ }\bibfield  {title} {\emph
  {\bibinfo {title} {Demonstration of self-seeding in a hard-x-ray
  free-electron laser},\ }}\href {\doibase 10.1038/nphoton.2012.180} {\bibfield
   {journal} {\bibinfo  {journal} {Nature Photonics}\ }\textbf {\bibinfo
  {volume} {6}},\ \bibinfo {pages} {693} (\bibinfo {year} {2012})}\BibitemShut
  {NoStop}%
\bibitem [{\citenamefont {Nam}\ \emph {et~al.}(2021)\citenamefont {Nam},
  \citenamefont {Min}, \citenamefont {Oh}, \citenamefont {Kim}, \citenamefont
  {Na}, \citenamefont {Suh}, \citenamefont {Yang}, \citenamefont {Cho},
  \citenamefont {Kim}, \citenamefont {Kim}, \citenamefont {Shim}, \citenamefont
  {Ko}, \citenamefont {Heo}, \citenamefont {Park}, \citenamefont {Kim},
  \citenamefont {Park}, \citenamefont {Park}, \citenamefont {Kim},
  \citenamefont {Chun}, \citenamefont {Hyun}, \citenamefont {Lee},
  \citenamefont {Kim}, \citenamefont {Eom}, \citenamefont {Rah}, \citenamefont
  {Shu}, \citenamefont {Kim}, \citenamefont {Terentyev}, \citenamefont {Blank},
  \citenamefont {Shvyd'ko}, \citenamefont {Lee},\ and\ \citenamefont
  {Kang}}]{Nam2021}%
  \BibitemOpen
  \bibfield  {author} {\bibinfo {author} {\bibfnamefont {I.}~\bibnamefont
  {Nam}}, \bibinfo {author} {\bibfnamefont {C.-K.}\ \bibnamefont {Min}},
  \bibinfo {author} {\bibfnamefont {B.}~\bibnamefont {Oh}}, \bibinfo {author}
  {\bibfnamefont {G.}~\bibnamefont {Kim}}, \bibinfo {author} {\bibfnamefont
  {D.}~\bibnamefont {Na}}, \bibinfo {author} {\bibfnamefont {Y.~J.}\
  \bibnamefont {Suh}}, \bibinfo {author} {\bibfnamefont {H.}~\bibnamefont
  {Yang}}, \bibinfo {author} {\bibfnamefont {M.~H.}\ \bibnamefont {Cho}},
  \bibinfo {author} {\bibfnamefont {C.}~\bibnamefont {Kim}}, \bibinfo {author}
  {\bibfnamefont {M.-J.}\ \bibnamefont {Kim}}, \bibinfo {author} {\bibfnamefont
  {C.~H.}\ \bibnamefont {Shim}}, \bibinfo {author} {\bibfnamefont {J.~H.}\
  \bibnamefont {Ko}}, \bibinfo {author} {\bibfnamefont {H.}~\bibnamefont
  {Heo}}, \bibinfo {author} {\bibfnamefont {J.}~\bibnamefont {Park}}, \bibinfo
  {author} {\bibfnamefont {J.}~\bibnamefont {Kim}}, \bibinfo {author}
  {\bibfnamefont {S.}~\bibnamefont {Park}}, \bibinfo {author} {\bibfnamefont
  {G.}~\bibnamefont {Park}}, \bibinfo {author} {\bibfnamefont {S.}~\bibnamefont
  {Kim}}, \bibinfo {author} {\bibfnamefont {S.~H.}\ \bibnamefont {Chun}},
  \bibinfo {author} {\bibfnamefont {H.}~\bibnamefont {Hyun}}, \bibinfo {author}
  {\bibfnamefont {J.~H.}\ \bibnamefont {Lee}}, \bibinfo {author} {\bibfnamefont
  {K.~S.}\ \bibnamefont {Kim}}, \bibinfo {author} {\bibfnamefont
  {I.}~\bibnamefont {Eom}}, \bibinfo {author} {\bibfnamefont {S.}~\bibnamefont
  {Rah}}, \bibinfo {author} {\bibfnamefont {D.}~\bibnamefont {Shu}}, \bibinfo
  {author} {\bibfnamefont {K.-J.}\ \bibnamefont {Kim}}, \bibinfo {author}
  {\bibfnamefont {S.}~\bibnamefont {Terentyev}}, \bibinfo {author}
  {\bibfnamefont {V.}~\bibnamefont {Blank}}, \bibinfo {author} {\bibfnamefont
  {Y.}~\bibnamefont {Shvyd'ko}}, \bibinfo {author} {\bibfnamefont {S.~J.}\
  \bibnamefont {Lee}}, \ and\ \bibinfo {author} {\bibfnamefont {H.-S.}\
  \bibnamefont {Kang}},\ }\bibfield  {title} {\emph {\bibinfo {title}
  {High-brightness self-seeded x-ray free-electron laser covering the
  3.5{\thinspace}kev to 14.6{\thinspace}kev range},\ }}\href {\doibase
  10.1038/s41566-021-00777-z} {\bibfield  {journal} {\bibinfo  {journal}
  {Nature Photonics}\ }\textbf {\bibinfo {volume} {15}},\ \bibinfo {pages}
  {435} (\bibinfo {year} {2021})}\BibitemShut {NoStop}%
\bibitem [{\citenamefont {Liu}\ \emph {et~al.}(2023)\citenamefont {Liu},
  \citenamefont {Grech}, \citenamefont {Guetg}, \citenamefont {Karabekyan},
  \citenamefont {Kocharyan}, \citenamefont {Kujala}, \citenamefont {Lechner},
  \citenamefont {Long}, \citenamefont {Mirian}, \citenamefont {Qin},
  \citenamefont {Serkez}, \citenamefont {Tomin}, \citenamefont {Yan},
  \citenamefont {Abeghyan}, \citenamefont {Anton}, \citenamefont {Blank},
  \citenamefont {Boesenberg}, \citenamefont {Brinker}, \citenamefont {Chen},
  \citenamefont {Decking}, \citenamefont {Dong}, \citenamefont {Kearney},
  \citenamefont {La~Civita}, \citenamefont {Madsen}, \citenamefont
  {Maltezopoulos}, \citenamefont {Rodriguez-Fernandez}, \citenamefont {Saldin},
  \citenamefont {Samoylova}, \citenamefont {Scholz}, \citenamefont {Sinn},
  \citenamefont {Sleziona}, \citenamefont {Shu}, \citenamefont {Tanikawa},
  \citenamefont {Terentiev}, \citenamefont {Trebushinin}, \citenamefont
  {Tschentscher}, \citenamefont {Vannoni}, \citenamefont {Wohlenberg},
  \citenamefont {Yakopov},\ and\ \citenamefont {Geloni}}]{Liu2023}%
  \BibitemOpen
  \bibfield  {author} {\bibinfo {author} {\bibfnamefont {S.}~\bibnamefont
  {Liu}}, \bibinfo {author} {\bibfnamefont {C.}~\bibnamefont {Grech}}, \bibinfo
  {author} {\bibfnamefont {M.}~\bibnamefont {Guetg}}, \bibinfo {author}
  {\bibfnamefont {S.}~\bibnamefont {Karabekyan}}, \bibinfo {author}
  {\bibfnamefont {V.}~\bibnamefont {Kocharyan}}, \bibinfo {author}
  {\bibfnamefont {N.}~\bibnamefont {Kujala}}, \bibinfo {author} {\bibfnamefont
  {C.}~\bibnamefont {Lechner}}, \bibinfo {author} {\bibfnamefont
  {T.}~\bibnamefont {Long}}, \bibinfo {author} {\bibfnamefont {N.}~\bibnamefont
  {Mirian}}, \bibinfo {author} {\bibfnamefont {W.}~\bibnamefont {Qin}},
  \bibinfo {author} {\bibfnamefont {S.}~\bibnamefont {Serkez}}, \bibinfo
  {author} {\bibfnamefont {S.}~\bibnamefont {Tomin}}, \bibinfo {author}
  {\bibfnamefont {J.}~\bibnamefont {Yan}}, \bibinfo {author} {\bibfnamefont
  {S.}~\bibnamefont {Abeghyan}}, \bibinfo {author} {\bibfnamefont
  {J.}~\bibnamefont {Anton}}, \bibinfo {author} {\bibfnamefont
  {V.}~\bibnamefont {Blank}}, \bibinfo {author} {\bibfnamefont
  {U.}~\bibnamefont {Boesenberg}}, \bibinfo {author} {\bibfnamefont
  {F.}~\bibnamefont {Brinker}}, \bibinfo {author} {\bibfnamefont
  {Y.}~\bibnamefont {Chen}}, \bibinfo {author} {\bibfnamefont {W.}~\bibnamefont
  {Decking}}, \bibinfo {author} {\bibfnamefont {X.}~\bibnamefont {Dong}},
  \bibinfo {author} {\bibfnamefont {S.}~\bibnamefont {Kearney}}, \bibinfo
  {author} {\bibfnamefont {D.}~\bibnamefont {La~Civita}}, \bibinfo {author}
  {\bibfnamefont {A.}~\bibnamefont {Madsen}}, \bibinfo {author} {\bibfnamefont
  {T.}~\bibnamefont {Maltezopoulos}}, \bibinfo {author} {\bibfnamefont
  {A.}~\bibnamefont {Rodriguez-Fernandez}}, \bibinfo {author} {\bibfnamefont
  {E.}~\bibnamefont {Saldin}}, \bibinfo {author} {\bibfnamefont
  {L.}~\bibnamefont {Samoylova}}, \bibinfo {author} {\bibfnamefont
  {M.}~\bibnamefont {Scholz}}, \bibinfo {author} {\bibfnamefont
  {H.}~\bibnamefont {Sinn}}, \bibinfo {author} {\bibfnamefont {V.}~\bibnamefont
  {Sleziona}}, \bibinfo {author} {\bibfnamefont {D.}~\bibnamefont {Shu}},
  \bibinfo {author} {\bibfnamefont {T.}~\bibnamefont {Tanikawa}}, \bibinfo
  {author} {\bibfnamefont {S.}~\bibnamefont {Terentiev}}, \bibinfo {author}
  {\bibfnamefont {A.}~\bibnamefont {Trebushinin}}, \bibinfo {author}
  {\bibfnamefont {T.}~\bibnamefont {Tschentscher}}, \bibinfo {author}
  {\bibfnamefont {M.}~\bibnamefont {Vannoni}}, \bibinfo {author} {\bibfnamefont
  {T.}~\bibnamefont {Wohlenberg}}, \bibinfo {author} {\bibfnamefont
  {M.}~\bibnamefont {Yakopov}}, \ and\ \bibinfo {author} {\bibfnamefont
  {G.}~\bibnamefont {Geloni}},\ }\bibfield  {title} {\emph {\bibinfo {title}
  {Cascaded hard x-ray self-seeded free-electron laser at megahertz repetition
  rate},\ }}\href {\doibase 10.1038/s41566-023-01305-x} {\bibfield  {journal}
  {\bibinfo  {journal} {Nature Photonics}\ }\textbf {\bibinfo {volume} {17}},\
  \bibinfo {pages} {984} (\bibinfo {year} {2023})}\BibitemShut {NoStop}%
\bibitem [{\citenamefont {McCall}\ and\ \citenamefont
  {Hahn}(1967)}]{McCall1967}%
  \BibitemOpen
  \bibfield  {author} {\bibinfo {author} {\bibfnamefont {S.~L.}\ \bibnamefont
  {McCall}}\ and\ \bibinfo {author} {\bibfnamefont {E.~L.}\ \bibnamefont
  {Hahn}},\ }\bibfield  {title} {\emph {\bibinfo {title} {Self-induced
  transparency by pulsed coherent light},\ }}\href {\doibase
  10.1103/PhysRevLett.18.908} {\bibfield  {journal} {\bibinfo  {journal} {Phys.
  Rev. Lett.}\ }\textbf {\bibinfo {volume} {18}},\ \bibinfo {pages} {908}
  (\bibinfo {year} {1967})}\BibitemShut {NoStop}%
\bibitem [{\citenamefont {Eberly}(1998)}]{Eberly1998}%
  \BibitemOpen
  \bibfield  {author} {\bibinfo {author} {\bibfnamefont {J.}~\bibnamefont
  {Eberly}},\ }\bibfield  {title} {\emph {\bibinfo {title} {Area theorem
  rederived},\ }}\href {\doibase 10.1364/OE.2.000173} {\bibfield  {journal}
  {\bibinfo  {journal} {Opt. Express}\ }\textbf {\bibinfo {volume} {2}},\
  \bibinfo {pages} {173} (\bibinfo {year} {1998})}\BibitemShut {NoStop}%
\bibitem [{\citenamefont {Shore}(2011)}]{Shore2011}%
  \BibitemOpen
  \bibfield  {author} {\bibinfo {author} {\bibfnamefont {B.~W.}\ \bibnamefont
  {Shore}},\ }\href {\doibase 10.1017/CBO9780511675713} {\emph {\bibinfo
  {title} {Manipulating Quantum Structures Using Laser Pulses}}}\ (\bibinfo
  {publisher} {Cambridge University Press},\ \bibinfo {year}
  {2011})\BibitemShut {NoStop}%
\bibitem [{\citenamefont {Allen}\ and\ \citenamefont
  {Eberly}(1987)}]{Allen1987_BOOK}%
  \BibitemOpen
  \bibfield  {author} {\bibinfo {author} {\bibfnamefont {L.}~\bibnamefont
  {Allen}}\ and\ \bibinfo {author} {\bibfnamefont {J.}~\bibnamefont {Eberly}},\
  }\href@noop {} {\emph {\bibinfo {title} {Optical Resonance and Two-level
  Atoms}}},\ Dover books on physics and chemistry\ (\bibinfo  {publisher}
  {Dover},\ \bibinfo {year} {1987})\BibitemShut {NoStop}%
\bibitem [{sup()}]{supplement_nonlin2024}%
  \BibitemOpen
  \href@noop {} {}\bibinfo {howpublished} {See Supplemental Material at XXXX
  for details on the numerical algorithm to compute the field enhancement of
  focused x-ray beams in thin-film cavities.}\BibitemShut {Stop}%
\bibitem [{\citenamefont {R\"ohlsberger}\ \emph {et~al.}(2002)\citenamefont
  {R\"ohlsberger}, \citenamefont {Thomas}, \citenamefont {Schlage},
  \citenamefont {Burkel}, \citenamefont {Leupold},\ and\ \citenamefont
  {R\"uffer}}]{Rohlsberger2002}%
  \BibitemOpen
  \bibfield  {author} {\bibinfo {author} {\bibfnamefont {R.}~\bibnamefont
  {R\"ohlsberger}}, \bibinfo {author} {\bibfnamefont {H.}~\bibnamefont
  {Thomas}}, \bibinfo {author} {\bibfnamefont {K.}~\bibnamefont {Schlage}},
  \bibinfo {author} {\bibfnamefont {E.}~\bibnamefont {Burkel}}, \bibinfo
  {author} {\bibfnamefont {O.}~\bibnamefont {Leupold}}, \ and\ \bibinfo
  {author} {\bibfnamefont {R.}~\bibnamefont {R\"uffer}},\ }\bibfield  {title}
  {\emph {\bibinfo {title} {Imaging the magnetic spin structure of
  exchange-coupled thin films},\ }}\href {\doibase
  10.1103/PhysRevLett.89.237201} {\bibfield  {journal} {\bibinfo  {journal}
  {Phys. Rev. Lett.}\ }\textbf {\bibinfo {volume} {89}},\ \bibinfo {pages}
  {237201} (\bibinfo {year} {2002})}\BibitemShut {NoStop}%
\bibitem [{\citenamefont {Tschentscher}\ \emph {et~al.}(2017)\citenamefont
  {Tschentscher}, \citenamefont {Bressler}, \citenamefont {Grünert},
  \citenamefont {Madsen}, \citenamefont {Mancuso}, \citenamefont {Meyer},
  \citenamefont {Scherz}, \citenamefont {Sinn},\ and\ \citenamefont
  {Zastrau}}]{Tschentscher2017}%
  \BibitemOpen
  \bibfield  {author} {\bibinfo {author} {\bibfnamefont {T.}~\bibnamefont
  {Tschentscher}}, \bibinfo {author} {\bibfnamefont {C.}~\bibnamefont
  {Bressler}}, \bibinfo {author} {\bibfnamefont {J.}~\bibnamefont {Grünert}},
  \bibinfo {author} {\bibfnamefont {A.}~\bibnamefont {Madsen}}, \bibinfo
  {author} {\bibfnamefont {A.~P.}\ \bibnamefont {Mancuso}}, \bibinfo {author}
  {\bibfnamefont {M.}~\bibnamefont {Meyer}}, \bibinfo {author} {\bibfnamefont
  {A.}~\bibnamefont {Scherz}}, \bibinfo {author} {\bibfnamefont
  {H.}~\bibnamefont {Sinn}}, \ and\ \bibinfo {author} {\bibfnamefont
  {U.}~\bibnamefont {Zastrau}},\ }\bibfield  {title} {\emph {\bibinfo {title}
  {Photon beam transport and scientific instruments at the european xfel},\
  }}\href {\doibase 10.3390/app7060592} {\bibfield  {journal} {\bibinfo
  {journal} {Applied Sciences}\ }\textbf {\bibinfo {volume} {7}} (\bibinfo
  {year} {2017}),\ 10.3390/app7060592}\BibitemShut {NoStop}%
\bibitem [{\citenamefont {Madsen}\ \emph {et~al.}(2021)\citenamefont {Madsen},
  \citenamefont {Hallmann}, \citenamefont {Ansaldi}, \citenamefont {Roth},
  \citenamefont {Lu}, \citenamefont {Kim}, \citenamefont {Boesenberg},
  \citenamefont {Zozulya}, \citenamefont {M{\"{o}}ller}, \citenamefont
  {Shayduk}, \citenamefont {Scholz}, \citenamefont {Bartmann}, \citenamefont
  {Schmidt}, \citenamefont {Lobato}, \citenamefont {Sukharnikov}, \citenamefont
  {Reiser}, \citenamefont {Kazarian},\ and\ \citenamefont
  {Petrov}}]{Madsen2021}%
  \BibitemOpen
  \bibfield  {author} {\bibinfo {author} {\bibfnamefont {A.}~\bibnamefont
  {Madsen}}, \bibinfo {author} {\bibfnamefont {J.}~\bibnamefont {Hallmann}},
  \bibinfo {author} {\bibfnamefont {G.}~\bibnamefont {Ansaldi}}, \bibinfo
  {author} {\bibfnamefont {T.}~\bibnamefont {Roth}}, \bibinfo {author}
  {\bibfnamefont {W.}~\bibnamefont {Lu}}, \bibinfo {author} {\bibfnamefont
  {C.}~\bibnamefont {Kim}}, \bibinfo {author} {\bibfnamefont {U.}~\bibnamefont
  {Boesenberg}}, \bibinfo {author} {\bibfnamefont {A.}~\bibnamefont {Zozulya}},
  \bibinfo {author} {\bibfnamefont {J.}~\bibnamefont {M{\"{o}}ller}}, \bibinfo
  {author} {\bibfnamefont {R.}~\bibnamefont {Shayduk}}, \bibinfo {author}
  {\bibfnamefont {M.}~\bibnamefont {Scholz}}, \bibinfo {author} {\bibfnamefont
  {A.}~\bibnamefont {Bartmann}}, \bibinfo {author} {\bibfnamefont
  {A.}~\bibnamefont {Schmidt}}, \bibinfo {author} {\bibfnamefont
  {I.}~\bibnamefont {Lobato}}, \bibinfo {author} {\bibfnamefont
  {K.}~\bibnamefont {Sukharnikov}}, \bibinfo {author} {\bibfnamefont
  {M.}~\bibnamefont {Reiser}}, \bibinfo {author} {\bibfnamefont
  {K.}~\bibnamefont {Kazarian}}, \ and\ \bibinfo {author} {\bibfnamefont
  {I.}~\bibnamefont {Petrov}},\ }\bibfield  {title} {\emph {\bibinfo {title}
  {{Materials Imaging and Dynamics (MID) instrument at the European X-ray
  Free-Electron Laser Facility}},\ }}\href {\doibase 10.1107/S1600577521001302}
  {\bibfield  {journal} {\bibinfo  {journal} {Journal of Synchrotron
  Radiation}\ }\textbf {\bibinfo {volume} {28}},\ \bibinfo {pages} {637}
  (\bibinfo {year} {2021})}\BibitemShut {NoStop}%
\bibitem [{\citenamefont {Madsen}\ \emph {et~al.}(2013)\citenamefont {Madsen},
  \citenamefont {Hallmann}, \citenamefont {Roth},\ and\ \citenamefont
  {Ansaldi}}]{Madsen2013_designRep}%
  \BibitemOpen
  \bibfield  {author} {\bibinfo {author} {\bibfnamefont {A.}~\bibnamefont
  {Madsen}}, \bibinfo {author} {\bibfnamefont {J.}~\bibnamefont {Hallmann}},
  \bibinfo {author} {\bibfnamefont {T.}~\bibnamefont {Roth}}, \ and\ \bibinfo
  {author} {\bibfnamefont {G.}~\bibnamefont {Ansaldi}},\ }\href@noop {}
  {\bibinfo {title} {Technical design report - scientific instrument mid},\
  }\bibinfo {howpublished}
  {\url{https://www.xfel.eu/sites/sites_custom/site_xfel/content/e35165/e46561/e46883/e46942/e46945/xfel_file46946/TR-2013-005_TDR_MID_eng.pdf}}
  (\bibinfo {year} {2013}),\ \bibinfo {note} {accessed: 2021-07-25}\BibitemShut
  {NoStop}%
\bibitem [{\citenamefont {Nakatsutsumi}\ \emph {et~al.}(2014)\citenamefont
  {Nakatsutsumi}, \citenamefont {Appel}, \citenamefont {Priebe}, \citenamefont
  {Thorpe}, \citenamefont {Pelka}, \citenamefont {Muller},\ and\ \citenamefont
  {Tschentscher}}]{Nakatsutsumi2014_designRep}%
  \BibitemOpen
  \bibfield  {author} {\bibinfo {author} {\bibfnamefont {M.}~\bibnamefont
  {Nakatsutsumi}}, \bibinfo {author} {\bibfnamefont {K.}~\bibnamefont {Appel}},
  \bibinfo {author} {\bibfnamefont {G.}~\bibnamefont {Priebe}}, \bibinfo
  {author} {\bibfnamefont {I.}~\bibnamefont {Thorpe}}, \bibinfo {author}
  {\bibfnamefont {A.}~\bibnamefont {Pelka}}, \bibinfo {author} {\bibfnamefont
  {B.}~\bibnamefont {Muller}}, \ and\ \bibinfo {author} {\bibfnamefont
  {T.}~\bibnamefont {Tschentscher}},\ }\href@noop {} {\bibinfo {title}
  {Technical design report - scientific instrument hed},\ }\bibinfo
  {howpublished}
  {\url{https://www.xfel.eu/sites/sites_custom/site_xfel/content/e35165/e46561/e46886/e46954/e46959/xfel_file46960/TR-2014-001_TDR_HED_eng.pdf}}
  (\bibinfo {year} {2014}),\ \bibinfo {note} {accessed: 2021-07-25}\BibitemShut
  {NoStop}%
\bibitem [{\citenamefont {Adams}\ \emph {et~al.}(2019)\citenamefont {Adams},
  \citenamefont {Aeppli}, \citenamefont {Allison}, \citenamefont {Baron},
  \citenamefont {Bucksbaum}, \citenamefont {Chumakov}, \citenamefont {Corder},
  \citenamefont {Cramer}, \citenamefont {DeBeer}, \citenamefont {Ding},
  \citenamefont {Evers}, \citenamefont {Frisch}, \citenamefont {Fuchs},
  \citenamefont {Gr\"ubel}, \citenamefont {Hastings}, \citenamefont {Heyl},
  \citenamefont {Holberg}, \citenamefont {Huang}, \citenamefont {Ishikawa},
  \citenamefont {Kaldun}, \citenamefont {Kim}, \citenamefont {Kolodziej},
  \citenamefont {Krzywinski}, \citenamefont {Li}, \citenamefont {Liao},
  \citenamefont {Lindberg}, \citenamefont {Madsen}, \citenamefont {Maxwell},
  \citenamefont {Monaco}, \citenamefont {Nelson}, \citenamefont {Palffy},
  \citenamefont {Porat}, \citenamefont {Qin}, \citenamefont {Raubenheimer},
  \citenamefont {Reis}, \citenamefont {R\"ohlsberger}, \citenamefont {Santra},
  \citenamefont {Schoenlein}, \citenamefont {Sch\"unemann}, \citenamefont
  {Shpyrko}, \citenamefont {Shvyd'ko}, \citenamefont {Shwartz}, \citenamefont
  {Singer}, \citenamefont {Sinha}, \citenamefont {Sutton}, \citenamefont
  {Tamasaku}, \citenamefont {Wille}, \citenamefont {Yabashi}, \citenamefont
  {Ye},\ and\ \citenamefont {Zhu}}]{Adams2019}%
  \BibitemOpen
  \bibfield  {author} {\bibinfo {author} {\bibfnamefont {B.}~\bibnamefont
  {Adams}}, \bibinfo {author} {\bibfnamefont {G.}~\bibnamefont {Aeppli}},
  \bibinfo {author} {\bibfnamefont {T.}~\bibnamefont {Allison}}, \bibinfo
  {author} {\bibfnamefont {A.~Q.~R.}\ \bibnamefont {Baron}}, \bibinfo {author}
  {\bibfnamefont {P.}~\bibnamefont {Bucksbaum}}, \bibinfo {author}
  {\bibfnamefont {A.~I.}\ \bibnamefont {Chumakov}}, \bibinfo {author}
  {\bibfnamefont {C.}~\bibnamefont {Corder}}, \bibinfo {author} {\bibfnamefont
  {S.~P.}\ \bibnamefont {Cramer}}, \bibinfo {author} {\bibfnamefont
  {S.}~\bibnamefont {DeBeer}}, \bibinfo {author} {\bibfnamefont
  {Y.}~\bibnamefont {Ding}}, \bibinfo {author} {\bibfnamefont {J.}~\bibnamefont
  {Evers}}, \bibinfo {author} {\bibfnamefont {J.}~\bibnamefont {Frisch}},
  \bibinfo {author} {\bibfnamefont {M.}~\bibnamefont {Fuchs}}, \bibinfo
  {author} {\bibfnamefont {G.}~\bibnamefont {Gr\"ubel}}, \bibinfo {author}
  {\bibfnamefont {J.~B.}\ \bibnamefont {Hastings}}, \bibinfo {author}
  {\bibfnamefont {C.~M.}\ \bibnamefont {Heyl}}, \bibinfo {author}
  {\bibfnamefont {L.}~\bibnamefont {Holberg}}, \bibinfo {author} {\bibfnamefont
  {Z.}~\bibnamefont {Huang}}, \bibinfo {author} {\bibfnamefont
  {T.}~\bibnamefont {Ishikawa}}, \bibinfo {author} {\bibfnamefont
  {A.}~\bibnamefont {Kaldun}}, \bibinfo {author} {\bibfnamefont {K.-J.}\
  \bibnamefont {Kim}}, \bibinfo {author} {\bibfnamefont {T.}~\bibnamefont
  {Kolodziej}}, \bibinfo {author} {\bibfnamefont {J.}~\bibnamefont
  {Krzywinski}}, \bibinfo {author} {\bibfnamefont {Z.}~\bibnamefont {Li}},
  \bibinfo {author} {\bibfnamefont {W.-T.}\ \bibnamefont {Liao}}, \bibinfo
  {author} {\bibfnamefont {R.}~\bibnamefont {Lindberg}}, \bibinfo {author}
  {\bibfnamefont {A.}~\bibnamefont {Madsen}}, \bibinfo {author} {\bibfnamefont
  {T.}~\bibnamefont {Maxwell}}, \bibinfo {author} {\bibfnamefont
  {G.}~\bibnamefont {Monaco}}, \bibinfo {author} {\bibfnamefont
  {K.}~\bibnamefont {Nelson}}, \bibinfo {author} {\bibfnamefont
  {A.}~\bibnamefont {Palffy}}, \bibinfo {author} {\bibfnamefont
  {G.}~\bibnamefont {Porat}}, \bibinfo {author} {\bibfnamefont
  {W.}~\bibnamefont {Qin}}, \bibinfo {author} {\bibfnamefont {T.}~\bibnamefont
  {Raubenheimer}}, \bibinfo {author} {\bibfnamefont {D.~A.}\ \bibnamefont
  {Reis}}, \bibinfo {author} {\bibfnamefont {R.}~\bibnamefont {R\"ohlsberger}},
  \bibinfo {author} {\bibfnamefont {R.}~\bibnamefont {Santra}}, \bibinfo
  {author} {\bibfnamefont {R.}~\bibnamefont {Schoenlein}}, \bibinfo {author}
  {\bibfnamefont {V.}~\bibnamefont {Sch\"unemann}}, \bibinfo {author}
  {\bibfnamefont {O.}~\bibnamefont {Shpyrko}}, \bibinfo {author} {\bibfnamefont
  {Y.}~\bibnamefont {Shvyd'ko}}, \bibinfo {author} {\bibfnamefont
  {S.}~\bibnamefont {Shwartz}}, \bibinfo {author} {\bibfnamefont
  {A.}~\bibnamefont {Singer}}, \bibinfo {author} {\bibfnamefont {S.~K.}\
  \bibnamefont {Sinha}}, \bibinfo {author} {\bibfnamefont {M.}~\bibnamefont
  {Sutton}}, \bibinfo {author} {\bibfnamefont {K.}~\bibnamefont {Tamasaku}},
  \bibinfo {author} {\bibfnamefont {H.-C.}\ \bibnamefont {Wille}}, \bibinfo
  {author} {\bibfnamefont {M.}~\bibnamefont {Yabashi}}, \bibinfo {author}
  {\bibfnamefont {J.}~\bibnamefont {Ye}}, \ and\ \bibinfo {author}
  {\bibfnamefont {D.}~\bibnamefont {Zhu}},\ }\href {\doibase
  10.48550/arXiv.1903.09317} {\bibinfo {title} {Scientific opportunities with
  an x-ray free-electron laser oscillator},\ } (\bibinfo {year} {2019}),\
  \Eprint {http://arxiv.org/abs/1903.09317} {arXiv:1903.09317
  [physics.ins-det]} \BibitemShut {NoStop}%
\bibitem [{web()}]{website_MID_calls}%
  \BibitemOpen
  \href {https://www.xfel.eu/facility/instruments/mid/index_eng.html} {\bibinfo
  {title} {Scientific instrument mid},\ }\bibinfo {note} {[Online; accessed
  02-April-2024]}\BibitemShut {NoStop}%
\bibitem [{\citenamefont {Follath}\ \emph {et~al.}(2019)\citenamefont
  {Follath}, \citenamefont {Koyama}, \citenamefont {Lipp}, \citenamefont
  {Medvedev}, \citenamefont {Tono}, \citenamefont {Ohashi}, \citenamefont
  {Patthey}, \citenamefont {Yabashi},\ and\ \citenamefont
  {Ziaja}}]{Follath2019}%
  \BibitemOpen
  \bibfield  {author} {\bibinfo {author} {\bibfnamefont {R.}~\bibnamefont
  {Follath}}, \bibinfo {author} {\bibfnamefont {T.}~\bibnamefont {Koyama}},
  \bibinfo {author} {\bibfnamefont {V.}~\bibnamefont {Lipp}}, \bibinfo {author}
  {\bibfnamefont {N.}~\bibnamefont {Medvedev}}, \bibinfo {author}
  {\bibfnamefont {K.}~\bibnamefont {Tono}}, \bibinfo {author} {\bibfnamefont
  {H.}~\bibnamefont {Ohashi}}, \bibinfo {author} {\bibfnamefont
  {L.}~\bibnamefont {Patthey}}, \bibinfo {author} {\bibfnamefont
  {M.}~\bibnamefont {Yabashi}}, \ and\ \bibinfo {author} {\bibfnamefont
  {B.}~\bibnamefont {Ziaja}},\ }\bibfield  {title} {\emph {\bibinfo {title}
  {X-ray induced damage of b4c-coated bilayer materials under various
  irradiation conditions},\ }}\href {\doibase 10.1038/s41598-019-38556-0}
  {\bibfield  {journal} {\bibinfo  {journal} {Scientific Reports}\ }\textbf
  {\bibinfo {volume} {9}},\ \bibinfo {pages} {2029} (\bibinfo {year}
  {2019})}\BibitemShut {NoStop}%
\bibitem [{\citenamefont {Kim}\ \emph {et~al.}(2015)\citenamefont {Kim},
  \citenamefont {Nagahira}, \citenamefont {Koyama}, \citenamefont {Matsuyama},
  \citenamefont {Sano}, \citenamefont {Yabashi}, \citenamefont {Ohashi},
  \citenamefont {Ishikawa},\ and\ \citenamefont {Yamauchi}}]{Kim2015}%
  \BibitemOpen
  \bibfield  {author} {\bibinfo {author} {\bibfnamefont {J.}~\bibnamefont
  {Kim}}, \bibinfo {author} {\bibfnamefont {A.}~\bibnamefont {Nagahira}},
  \bibinfo {author} {\bibfnamefont {T.}~\bibnamefont {Koyama}}, \bibinfo
  {author} {\bibfnamefont {S.}~\bibnamefont {Matsuyama}}, \bibinfo {author}
  {\bibfnamefont {Y.}~\bibnamefont {Sano}}, \bibinfo {author} {\bibfnamefont
  {M.}~\bibnamefont {Yabashi}}, \bibinfo {author} {\bibfnamefont
  {H.}~\bibnamefont {Ohashi}}, \bibinfo {author} {\bibfnamefont
  {T.}~\bibnamefont {Ishikawa}}, \ and\ \bibinfo {author} {\bibfnamefont
  {K.}~\bibnamefont {Yamauchi}},\ }\bibfield  {title} {\emph {\bibinfo {title}
  {Damage threshold of platinum/carbon multilayers under hard x-ray
  free-electron laser irradiation},\ }}\href {\doibase 10.1364/OE.23.029032}
  {\bibfield  {journal} {\bibinfo  {journal} {Opt. Express}\ }\textbf {\bibinfo
  {volume} {23}},\ \bibinfo {pages} {29032} (\bibinfo {year}
  {2015})}\BibitemShut {NoStop}%
\bibitem [{\citenamefont {Lentrodt}(2021)}]{LentrodtPhD}%
  \BibitemOpen
  \bibfield  {author} {\bibinfo {author} {\bibfnamefont {D.}~\bibnamefont
  {Lentrodt}},\ }\href {\doibase 10.11588/heidok.00030671} {\bibinfo {title}
  {Ab initio approaches to x-ray cavity qed},\ } (\bibinfo {year}
  {2021})\BibitemShut {NoStop}%
\bibitem [{\citenamefont {Harris}\ \emph {et~al.}(2020)\citenamefont {Harris},
  \citenamefont {Millman}, \citenamefont {van~der Walt}, \citenamefont
  {Gommers}, \citenamefont {Virtanen}, \citenamefont {Cournapeau},
  \citenamefont {Wieser}, \citenamefont {Taylor}, \citenamefont {Berg},
  \citenamefont {Smith}, \citenamefont {Kern}, \citenamefont {Picus},
  \citenamefont {Hoyer}, \citenamefont {van Kerkwijk}, \citenamefont {Brett},
  \citenamefont {Haldane}, \citenamefont {Fern{\'a}ndez~del R{\'i}o},
  \citenamefont {Wiebe}, \citenamefont {Peterson}, \citenamefont
  {G{\'e}rard-Marchant}, \citenamefont {Sheppard}, \citenamefont {Reddy},
  \citenamefont {Weckesser}, \citenamefont {Abbasi}, \citenamefont {Gohlke},\
  and\ \citenamefont {Oliphant}}]{numpy2020}%
  \BibitemOpen
  \bibfield  {author} {\bibinfo {author} {\bibfnamefont {C.~R.}\ \bibnamefont
  {Harris}}, \bibinfo {author} {\bibfnamefont {K.~J.}\ \bibnamefont {Millman}},
  \bibinfo {author} {\bibfnamefont {S.~J.}\ \bibnamefont {van~der Walt}},
  \bibinfo {author} {\bibfnamefont {R.}~\bibnamefont {Gommers}}, \bibinfo
  {author} {\bibfnamefont {P.}~\bibnamefont {Virtanen}}, \bibinfo {author}
  {\bibfnamefont {D.}~\bibnamefont {Cournapeau}}, \bibinfo {author}
  {\bibfnamefont {E.}~\bibnamefont {Wieser}}, \bibinfo {author} {\bibfnamefont
  {J.}~\bibnamefont {Taylor}}, \bibinfo {author} {\bibfnamefont
  {S.}~\bibnamefont {Berg}}, \bibinfo {author} {\bibfnamefont {N.~J.}\
  \bibnamefont {Smith}}, \bibinfo {author} {\bibfnamefont {R.}~\bibnamefont
  {Kern}}, \bibinfo {author} {\bibfnamefont {M.}~\bibnamefont {Picus}},
  \bibinfo {author} {\bibfnamefont {S.}~\bibnamefont {Hoyer}}, \bibinfo
  {author} {\bibfnamefont {M.~H.}\ \bibnamefont {van Kerkwijk}}, \bibinfo
  {author} {\bibfnamefont {M.}~\bibnamefont {Brett}}, \bibinfo {author}
  {\bibfnamefont {A.}~\bibnamefont {Haldane}}, \bibinfo {author} {\bibfnamefont
  {J.}~\bibnamefont {Fern{\'a}ndez~del R{\'i}o}}, \bibinfo {author}
  {\bibfnamefont {M.}~\bibnamefont {Wiebe}}, \bibinfo {author} {\bibfnamefont
  {P.}~\bibnamefont {Peterson}}, \bibinfo {author} {\bibfnamefont
  {P.}~\bibnamefont {G{\'e}rard-Marchant}}, \bibinfo {author} {\bibfnamefont
  {K.}~\bibnamefont {Sheppard}}, \bibinfo {author} {\bibfnamefont
  {T.}~\bibnamefont {Reddy}}, \bibinfo {author} {\bibfnamefont
  {W.}~\bibnamefont {Weckesser}}, \bibinfo {author} {\bibfnamefont
  {H.}~\bibnamefont {Abbasi}}, \bibinfo {author} {\bibfnamefont
  {C.}~\bibnamefont {Gohlke}}, \ and\ \bibinfo {author} {\bibfnamefont {T.~E.}\
  \bibnamefont {Oliphant}},\ }\bibfield  {title} {\emph {\bibinfo {title}
  {Array programming with {NumPy}},\ }}\href {\doibase
  10.1038/s41586-020-2649-2} {\bibfield  {journal} {\bibinfo  {journal}
  {Nature}\ }\textbf {\bibinfo {volume} {585}},\ \bibinfo {pages} {357–362}
  (\bibinfo {year} {2020})}\BibitemShut {NoStop}%
\bibitem [{\citenamefont {Virtanen}\ \emph {et~al.}(2020)\citenamefont
  {Virtanen}, \citenamefont {Gommers}, \citenamefont {Oliphant}, \citenamefont
  {Haberland}, \citenamefont {Reddy}, \citenamefont {Cournapeau}, \citenamefont
  {Burovski}, \citenamefont {Peterson}, \citenamefont {Weckesser},
  \citenamefont {Bright}, \citenamefont {{van der Walt}}, \citenamefont
  {Brett}, \citenamefont {Wilson}, \citenamefont {Millman}, \citenamefont
  {Mayorov}, \citenamefont {Nelson}, \citenamefont {Jones}, \citenamefont
  {Kern}, \citenamefont {Larson}, \citenamefont {Carey}, \citenamefont {Polat},
  \citenamefont {Feng}, \citenamefont {Moore}, \citenamefont {{VanderPlas}},
  \citenamefont {Laxalde}, \citenamefont {Perktold}, \citenamefont {Cimrman},
  \citenamefont {Henriksen}, \citenamefont {Quintero}, \citenamefont {Harris},
  \citenamefont {Archibald}, \citenamefont {Ribeiro}, \citenamefont
  {Pedregosa}, \citenamefont {{van Mulbregt}},\ and\ \citenamefont {{SciPy 1.0
  Contributors}}}]{scipy2020}%
  \BibitemOpen
  \bibfield  {author} {\bibinfo {author} {\bibfnamefont {P.}~\bibnamefont
  {Virtanen}}, \bibinfo {author} {\bibfnamefont {R.}~\bibnamefont {Gommers}},
  \bibinfo {author} {\bibfnamefont {T.~E.}\ \bibnamefont {Oliphant}}, \bibinfo
  {author} {\bibfnamefont {M.}~\bibnamefont {Haberland}}, \bibinfo {author}
  {\bibfnamefont {T.}~\bibnamefont {Reddy}}, \bibinfo {author} {\bibfnamefont
  {D.}~\bibnamefont {Cournapeau}}, \bibinfo {author} {\bibfnamefont
  {E.}~\bibnamefont {Burovski}}, \bibinfo {author} {\bibfnamefont
  {P.}~\bibnamefont {Peterson}}, \bibinfo {author} {\bibfnamefont
  {W.}~\bibnamefont {Weckesser}}, \bibinfo {author} {\bibfnamefont
  {J.}~\bibnamefont {Bright}}, \bibinfo {author} {\bibfnamefont {S.~J.}\
  \bibnamefont {{van der Walt}}}, \bibinfo {author} {\bibfnamefont
  {M.}~\bibnamefont {Brett}}, \bibinfo {author} {\bibfnamefont
  {J.}~\bibnamefont {Wilson}}, \bibinfo {author} {\bibfnamefont {K.~J.}\
  \bibnamefont {Millman}}, \bibinfo {author} {\bibfnamefont {N.}~\bibnamefont
  {Mayorov}}, \bibinfo {author} {\bibfnamefont {A.~R.~J.}\ \bibnamefont
  {Nelson}}, \bibinfo {author} {\bibfnamefont {E.}~\bibnamefont {Jones}},
  \bibinfo {author} {\bibfnamefont {R.}~\bibnamefont {Kern}}, \bibinfo {author}
  {\bibfnamefont {E.}~\bibnamefont {Larson}}, \bibinfo {author} {\bibfnamefont
  {C.~J.}\ \bibnamefont {Carey}}, \bibinfo {author} {\bibfnamefont
  {{\.I}.}~\bibnamefont {Polat}}, \bibinfo {author} {\bibfnamefont
  {Y.}~\bibnamefont {Feng}}, \bibinfo {author} {\bibfnamefont {E.~W.}\
  \bibnamefont {Moore}}, \bibinfo {author} {\bibfnamefont {J.}~\bibnamefont
  {{VanderPlas}}}, \bibinfo {author} {\bibfnamefont {D.}~\bibnamefont
  {Laxalde}}, \bibinfo {author} {\bibfnamefont {J.}~\bibnamefont {Perktold}},
  \bibinfo {author} {\bibfnamefont {R.}~\bibnamefont {Cimrman}}, \bibinfo
  {author} {\bibfnamefont {I.}~\bibnamefont {Henriksen}}, \bibinfo {author}
  {\bibfnamefont {E.~A.}\ \bibnamefont {Quintero}}, \bibinfo {author}
  {\bibfnamefont {C.~R.}\ \bibnamefont {Harris}}, \bibinfo {author}
  {\bibfnamefont {A.~M.}\ \bibnamefont {Archibald}}, \bibinfo {author}
  {\bibfnamefont {A.~H.}\ \bibnamefont {Ribeiro}}, \bibinfo {author}
  {\bibfnamefont {F.}~\bibnamefont {Pedregosa}}, \bibinfo {author}
  {\bibfnamefont {P.}~\bibnamefont {{van Mulbregt}}}, \ and\ \bibinfo {author}
  {\bibnamefont {{SciPy 1.0 Contributors}}},\ }\bibfield  {title} {\emph
  {\bibinfo {title} {{{SciPy} 1.0: Fundamental Algorithms for Scientific
  Computing in Python}},\ }}\href {\doibase 10.1038/s41592-019-0686-2}
  {\bibfield  {journal} {\bibinfo  {journal} {Nature Methods}\ }\textbf
  {\bibinfo {volume} {17}},\ \bibinfo {pages} {261} (\bibinfo {year}
  {2020})}\BibitemShut {NoStop}%
\bibitem [{\citenamefont {Hunter}(2007)}]{matplotlib2007}%
  \BibitemOpen
  \bibfield  {author} {\bibinfo {author} {\bibfnamefont {J.~D.}\ \bibnamefont
  {Hunter}},\ }\bibfield  {title} {\emph {\bibinfo {title} {Matplotlib: A 2d
  graphics environment},\ }}\href {\doibase 10.1109/MCSE.2007.55} {\bibfield
  {journal} {\bibinfo  {journal} {Computing in Science \& Engineering}\
  }\textbf {\bibinfo {volume} {9}},\ \bibinfo {pages} {90} (\bibinfo {year}
  {2007})}\BibitemShut {NoStop}%
\end{thebibliography}%
	
\end{document}